\documentclass[12pt]{article}
\usepackage{fullpage}
\usepackage{amssymb, amsmath, amsthm}
\usepackage{graphicx}
\usepackage{booktabs}
\usepackage{multirow}
\newcommand{\argmax}{\operatorname*{arg \ max}}
\newcommand{\argmin}{\operatorname*{arg \ min}}

\usepackage{algorithm}
\usepackage{mathtools}
\usepackage{setspace}

\usepackage{breqn}
\usepackage{natbib}
\usepackage{fullpage}
\usepackage{hyperref}
\usepackage{tikz}
\usepackage{subfigure}
\usepackage{chngcntr}
\usepackage{array}
\usepackage[section]{placeins}
\newcolumntype{C}[1]{>{\centering\arraybackslash}p{#1}}
\usetikzlibrary{shapes.geometric}
\usetikzlibrary{arrows,automata}
\usepackage{amsmath}
\usepackage{soul}
\usepackage{bbold}
\theoremstyle{plain}

\newtheorem{lemma}{Lemma}

\newcommand{\mbalpha}{\boldsymbol{\alpha}}
\newcommand{\mbbeta}{\boldsymbol{\beta}}
\newcommand{\mbgamma}{\boldsymbol{\gamma}}
\newcommand{\mbphi}{\boldsymbol{\phi}}
\newcommand{\mbu}{\boldsymbol{u}}
\newcommand{\mbU}{\boldsymbol{U}}
\newcommand{\mbx}{\boldsymbol{x}}
\newcommand{\mbz}{\boldsymbol{z}}
\newcommand{\mbX}{\boldsymbol{X}}
\newcommand{\mbZ}{\boldsymbol{Z}}

\newcommand{\mbnu}{\boldsymbol{\nu}}
\newcommand{\mbP}{\boldsymbol{P}}
\newcommand{\mbC}{\boldsymbol{C}}
\newcommand{\mbI}{\boldsymbol{I}}

\makeatletter
\newcommand*{\centerfloat}{%
  \parindent \z@
  \leftskip \z@ \@plus 1fil \@minus \textwidth
  \rightskip\leftskip
  \parfillskip \z@skip}
\makeatother

\mathtoolsset{showonlyrefs}

\tikzset{
  treenode/.style = {align=center, inner sep=0pt, text centered,
    font=\sffamily},
  arn_n/.style = {treenode, ellipse, black, draw=black,
    text width=3em, minimum height=2.3em},% arbre rouge noir, noeud noir
  arn_r/.style = {treenode, ellipse, red, draw=red, %fill=lightgray,
    text width=3em, minimum height=2.3em},% arbre rouge noir, noeud rouge
  arn_x/.style = {treenode, rectangle, draw=black,
    minimum width=0.5em, minimumc height=0.5em}% arbre rouge noir, nil
}

\title{Binned multinomial logistic regression for integrative \\cell type annotation}
\author{Keshav Motwani$^1$, Rhonda Bacher$^{2,3}$, and Aaron J. Molstad$^{1, 3}$\footnote{Correspondence: amolstad@ufl.edu}\\
Department of Statistics$^1$, Department of Biostatistics$^2$, \\
and Genetics Institute$^3$, University of Florida\\
}
\date{}

\begin{document}
\maketitle
\begin{abstract}
Categorizing individual cells into one of many known cell type categories, also known as cell type annotation, is a critical step in the analysis of single-cell genomics data. The current process of annotation is time-intensive and subjective, which has led to different studies describing cell types with labels of varying degrees of resolution. While supervised learning approaches have provided automated solutions to annotation, there remains a significant challenge in fitting a unified model for multiple datasets with inconsistent labels. 
In this article, we propose a new multinomial logistic regression estimator which can be used to model cell type probabilities by integrating multiple datasets with labels of varying resolution. To compute our estimator, we solve a nonconvex optimization problem using a blockwise proximal gradient descent algorithm. 
We show through simulation studies that our approach estimates cell type probabilities more accurately than competitors in a wide variety of scenarios. We apply our method to ten single-cell RNA-seq datasets and demonstrate its utility in predicting fine resolution cell type labels on unlabeled data as well as refining cell type labels on data with existing coarse resolution annotations. An R package implementing the method is available at \url{https://github.com/keshav-motwani/IBMR} and the collection of datasets we analyze is available at \url{https://github.com/keshav-motwani/AnnotatedPBMC}.\smallskip

\end{abstract}

%\newpage
%\doublespacing
\onehalfspacing
\section{Introduction}
\subsection{Overview}

%The analysis of RNA expression from individual cells using single-cell RNA sequencing is central to modern genomics. 
One of the first and most important tasks in the analysis of single-cell data is cell type annotation, where individual cells are categorized into one of many known cell type categories having well-characterized biological functions. The vast majority of studies perform annotation by first clustering cells based on their gene expression and then manually labeling the clusters based on upregulated marker genes within each cluster \citep{tabula2018single}. This is often time-intensive and arguably subjective, as the set of cell type labels used is inconsistent across studies: they vary based on scientific interests of the investigators, aims of the study, and availability of external data. In turn, a large number of automated methods have been developed to standardize the cell type annotation process, for example, see Table 1 of \citet{PASQUINI2021961} and references therein. 

The vast majority of the existing approaches for automated cell type annotation fit a classification model using a single training dataset (e.g., a dataset collected and annotated by a single investigator/lab), treating normalized gene expression as predictors. 
Cell types in a new (unannotated) dataset are then predicted according to the fitted model. In \citet{abdelaal2019comparison}, more than 20 such methods were benchmarked and shown to perform well in a variety of settings. However, these methods tended to perform poorly in terms of prediction across datasets (varying by batch, lab, or protocols) and in datasets with a large number of labels (i.e., high resolution cell type categories) \citep{abdelaal2019comparison}. Furthermore, a crucial choice for these methods is deciding which dataset should be used to train the model. Datasets can differ in numerous ways, but most relevant to the task we consider: they can have drastically different cell type labels and differ in the amount of detail provided by each label across datasets \citep{ma2021evaluation}. The existing annotation approaches are also limited to single training datasets or multiple datasets with consistent cell type labels. Here, we propose a novel approach for automated annotation that overcomes these limitations.

\begin{figure}[!t]
\begin{tabular}{C{.5\textwidth}C{.5\textwidth}}
%%%%%%%%%%%%%%%%%%%%%% 2 %%%%%%%%%%%%%%%%%%%%%%%%%%%%%%%%%%%
\subfigure [Dataset 1] {
  \resizebox{0.5\textwidth}{!}{%
\begin{tikzpicture}[->,>=stealth',level/.style={sibling distance = 5.5cm/#1,
  level distance = 1.7cm}] 
\node [arn_n] {T}
    child{ node [arn_r] {CD4+} 
            child{ node [arn_n] {memory} 
                        child{ node [arn_n] {effector} 
            }
            child{ node [arn_n] {central}
            }
            }
            child{ node [arn_n] {naive}
            }                            
    }
    child{ node [arn_r] {CD8+}
            child{ node [arn_n] {memory} 
            child{ node [arn_n] {effector} 
            }
            child{ node [arn_n] {central}
            }
            }
            child{ node [arn_n] {naive}
            }
    }
; 
\end{tikzpicture}
    }
} &
%%%%%%%%%%%%%%%%%%%%%% 3 %%%%%%%%%%%%%%%%%%%%%%%%%%%%%%%%%%%
\subfigure [Dataset 2] {
  \resizebox{0.5\textwidth}{!}{%
\begin{tikzpicture}[->,>=stealth',level/.style={sibling distance = 5.5cm/#1,
  level distance = 1.7cm}] 
\node [arn_n] {T}
    child{ node [arn_n] {CD4+} 
            child{ node [arn_n] {memory} 
                        child{ node [arn_r] {effector} 
            }
            child{ node [arn_r] {central}
            }
            }
            child{ node [arn_r] {naive}
            }                            
    }
    child{ node [arn_r] {CD8+}
            child{ node [arn_n] {memory} 
            child{ node [arn_n] {effector} 
            }
            child{ node [arn_n] {central}
            }
            }
            child{ node [arn_n] {naive}
            }
    }
; 
\end{tikzpicture}
    }
} \\
%%%%%%%%%%%%%%%%%%%%%% 1 %%%%%%%%%%%%%%%%%%%%%%%%%%%%%%%%%%%
\subfigure [Dataset 3] {
    \resizebox{0.5\textwidth}{!}{%
\begin{tikzpicture}[->,>=stealth',level/.style={sibling distance = 5.5cm/#1,
  level distance = 1.7cm}] 
\node [arn_n] {T}
    child{ node [arn_n] {CD4+} 
            child{ node [arn_r] {memory} 
                        child{ node [arn_n] {effector} 
            }
            child{ node [arn_n] {central}
            }
            }
            child{ node [arn_r] {naive}
            }                            
    }
    child{ node [arn_n] {CD8+}
            child{ node [arn_n] {memory} 
            child{ node [arn_r] {effector} 
            }
            child{ node [arn_r] {central}
            }
            }
            child{ node [arn_r] {naive}
            }
    }
; 
\end{tikzpicture}
    }
} &
%%%%%%%%%%%%%%%%%%%%%% 1 %%%%%%%%%%%%%%%%%%%%%%%%%%%%%%%%%%%
\subfigure [Bins] {
\includegraphics[width=0.45\textwidth]{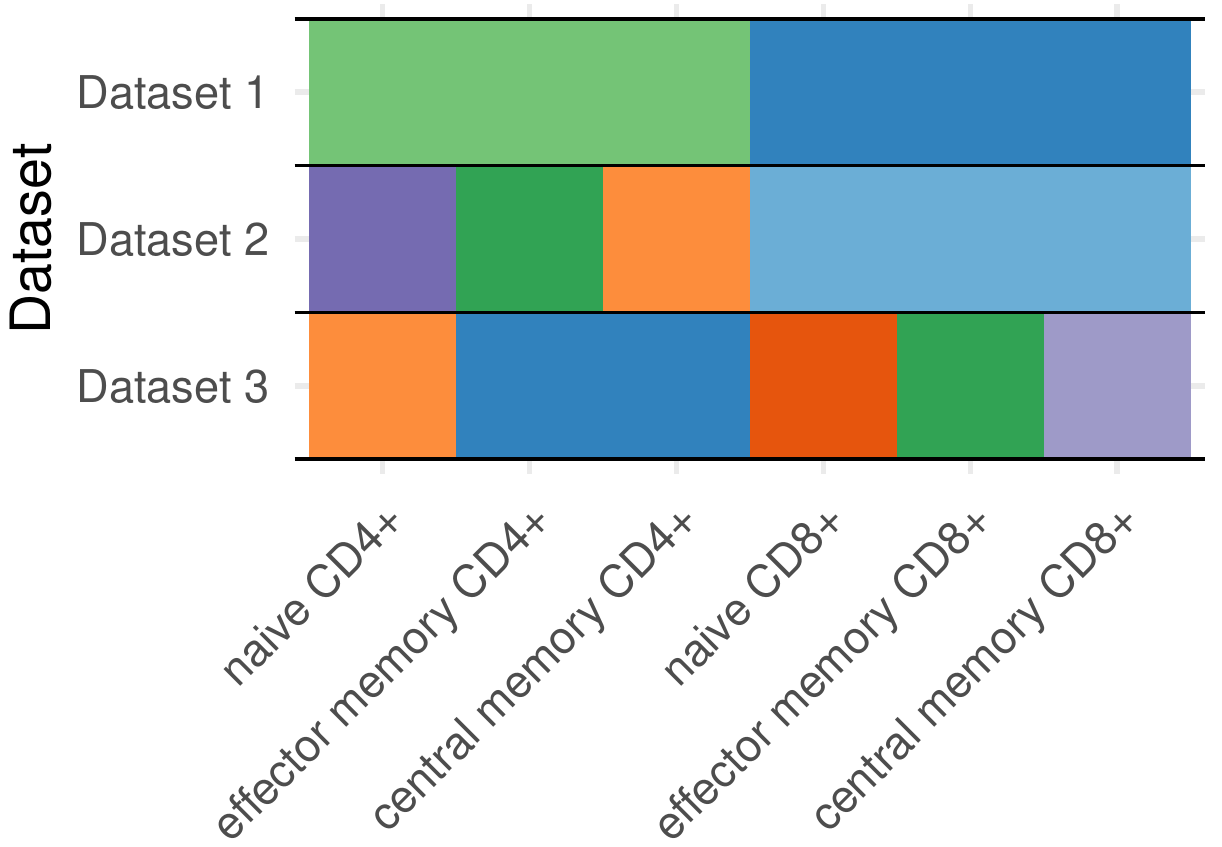}
} 
\end{tabular}
\caption{Illustrative example of label structure across datasets. (a-c) The tree depicts the true hierarchical structure of cell type categories, with cells in Datasets 1--3 annotated at different resolution labels, highlighted in red. Finest resolution categories are defined by the labels at the terminal nodes of the tree. (d) Graphical representation of binning functions for Datasets 1--3 (see Section \ref{subsec:binning_function}), where within each row, a unique color represents a label in that dataset which is a bin of finest resolution categories.\label{fig:varying_resolutions}}
\end{figure}
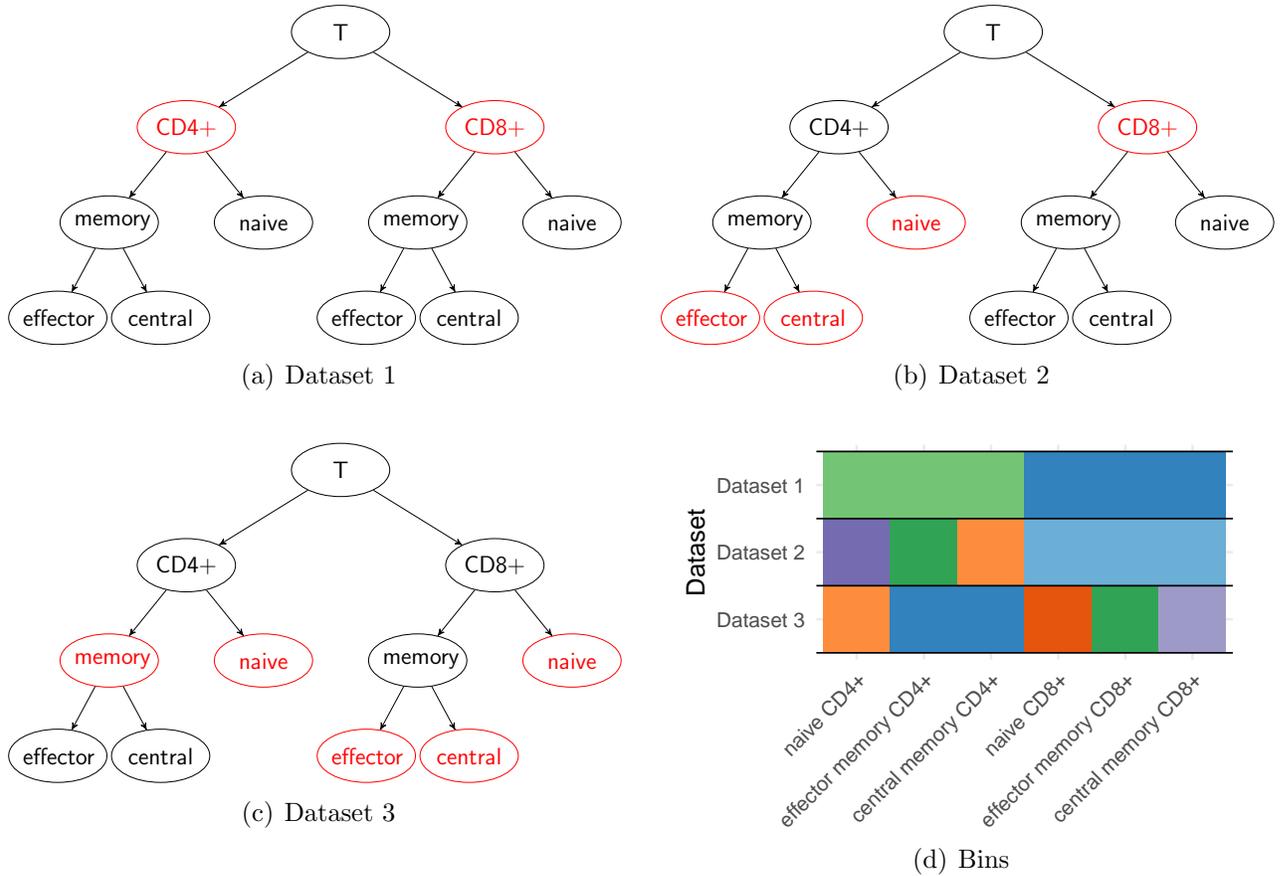

We begin by depicting the situation of differing degrees of resolution in labels used to annotate different datasets in Figure \ref{fig:varying_resolutions}. In this hypothetical situation, one has access to three datasets, Datasets 1, 2, and 3, each of which has been expertly annotated manually. In Dataset 1, cells are labeled as either CD4+ or CD8+. If one trained a model using only Dataset 1, the only possible predicted labels for a new dataset would be CD4+ or CD8+. In Dataset 2, the cells are labeled as one of naive CD4+, effector memory CD4+, central memory CD4+, or CD8+, so if one instead trained the model using Dataset 2, it would be possible to predict/annotate the subcategories of CD4+ T-cells with finer resolution labels when compared to Dataset 1. Dataset 3 has finer resolution labels for subcategories of CD8+ cells than Dataset 2, but does not distinguish between the two finer CD4+ memory cell types like Dataset 2. Thus, using a single dataset to train annotation models presents a trade-off between fine resolution labels for subcategories of CD4+ and subcategories of CD8+ cells. 

If one wanted to incorporate all datasets into the framework of existing annotation methods, the level of detail in the annotations of Dataset 2 and Dataset 3 must be reduced by labeling cells of all three datasets as one of CD4+ or CD8+. However, this results in a significant loss of information and may limit downstream scientific applications. Alternatively, one could mix-and-match subsets of cells from different datasets which have the most detail for specific cell types. In this example, it would mean taking the subcategories of CD4+ cells from Dataset 2 and subcategories of CD8+ cells from Dataset 3 and ignoring Dataset 1. As such, this approach would be less efficient than one which uses all available data, and moreover, will generalize poorly since technical differences across datasets (i.e., ``batch effects'') may be confounded with some cell type categories. Despite the existence of hundreds of publicly available datasets with expertly annotated cell types, existing methods are limited in their ability to integrate a wide-array of datasets due to varying label resolution. 

Ideally, we would like to use all the data from all three datasets to train an annotation model without any loss of information. 
% Of course, real single-cell datasets consist of more than one major cell type (T cells in the example from Figure \ref{fig:varying_resolutions}), and available labels do not follow the simplistic tree-structure across datasets as suggested in Figure \ref{fig:varying_resolutions}. For example, a fourth dataset may have an annotation called ``memory'' which includes cells belonging to the union of categories $\{$effector memory CD4+, central memory CD4+, effector memory CD8+, central memory CD8+$\}$. 
To do so, our proposed approach takes advantage of the ``binned'' label structures (Figure \ref{fig:varying_resolutions}d). For example, cells with the label CD4+ in Dataset 1, biologically, must belong to one of the following finest resolution categories: naive CD4+, effector memory CD4+, or central memory CD4+ The specific label, however, is unknown without additional analysis or manual annotation. In this article, we propose a new classification method which will allow investigators to (i) use all available datasets jointly to train a unified classification model without loss of information, and (ii) make cell type predictions/annotations at the finest resolution labels allowed by the union of all datasets' labels. For example, given the datasets depicted in Figure \ref{fig:varying_resolutions}, our method would fit a model using data from all cells from all three datasets, and would yield predicted probabilities of each cell belonging to the categories: naive CD4+, effector memory CD4+, central memory CD4+, naive CD8+, effector memory CD8+, or central memory CD8+ (i.e., the categories at the terminal nodes of the tree). Notably, our method does not require that labels are tree-structured as in this example. We require only that labels are amenable to ``binning'', which we describe in Section \ref{subsec:binning_function}.

\subsection{Existing approaches}

The issue of varying labels across datasets has been recognized in the recent single-cell literature. For example, \citet{shasha2021superscan} manually reannotated publicly available datasets which collected both single-cell gene expression and protein expression data, and fit a cell type classification model across all datasets using reannotated labels with extreme gradient boosting (XGBoost). To reannotate the data, they cleverly employed methods from the field of flow cytometry to ``gate'' cells based on protein expression using a series of bivariate protein expression plots and manually drawing shapes around groups of cells. This reannotation process, however, is very time-intensive and requires concurrent protein expression in cells. Even with this detailed approach, differences in protein measurements across datasets limited their ability to achieve consistently fine annotations across all datasets. Similarly, \citet{conde2021cross} employed a two-step reannotation process. First, with expert input, they attempted to reconcile and rename labels across datasets to achieve a consistent set of labels. Second, they fit a ridge-penalized multinomial logistic regression model on datasets for which they successfully renamed labels for, and used this model to predict the labels for the remaining unresolved datasets. Cells were clustered in each remaining dataset based on gene expression, and each cluster was labeled on a majority vote of the predictions for cells in that cluster. The predicted cluster labels were then treated as true labels for these datasets, and the model was refit using all of the datasets. This approach motivates a two-step approximation to our proposed method, which we term \texttt{relabel} (see Section \ref{subsec:comp_methods}) and compare to throughout this paper. 

% Mon Nov  1 14:03:22 2021
\begin{table}[t]
\centering
\begin{tabular}{lll}
  \toprule
  Dataset & \# of labels & Reference(s) \\ 
  \midrule
  \texttt{hao\_2020} & 28 & \citet{hao2020integrated} \\ 
   \texttt{tsang\_2021} & 18 & \citet{tsangliu2021time} \\ 
   \texttt{haniffa\_2021} & 16 & \citet{haniffastephenson2021single} \\ 
  \texttt{su\_2020} & 13 & \citet{su2020multi}, \citet{shasha2021superscan} \\ 
   \texttt{10x\_pbmc\_5k\_v3} & 12 & \citet{10x_pbmc_5k_v3}, \citet{shasha2021superscan} \\
    \texttt{blish\_2020} & 12 & \citet{blishwilk2020single} \\ 
  \texttt{kotliarov\_2020} & 9 & \citet{kotliarov2020broad} \\ 
   \texttt{10x\_pbmc\_10k} & 9 & \citet{10x_pbmc_10k}, \citet{shasha2021superscan} \\ 
  \texttt{10x\_sorted} & 8 & \citet{zheng2017massively} \\ 
   \texttt{ding\_2019} & 8 & \citet{ding2019systematic} \\ 
   \bottomrule
\end{tabular}
\caption{Number of labels and reference(s) for each of the peripheral blood single-cell genomics datasets analyzed in Section \ref{sec:data_analysis}.} \label{table:datasets}
\end{table}
\subsection{Motivating application}
\begin{figure}[t]
\begin{center}
\includegraphics[width=\textwidth]{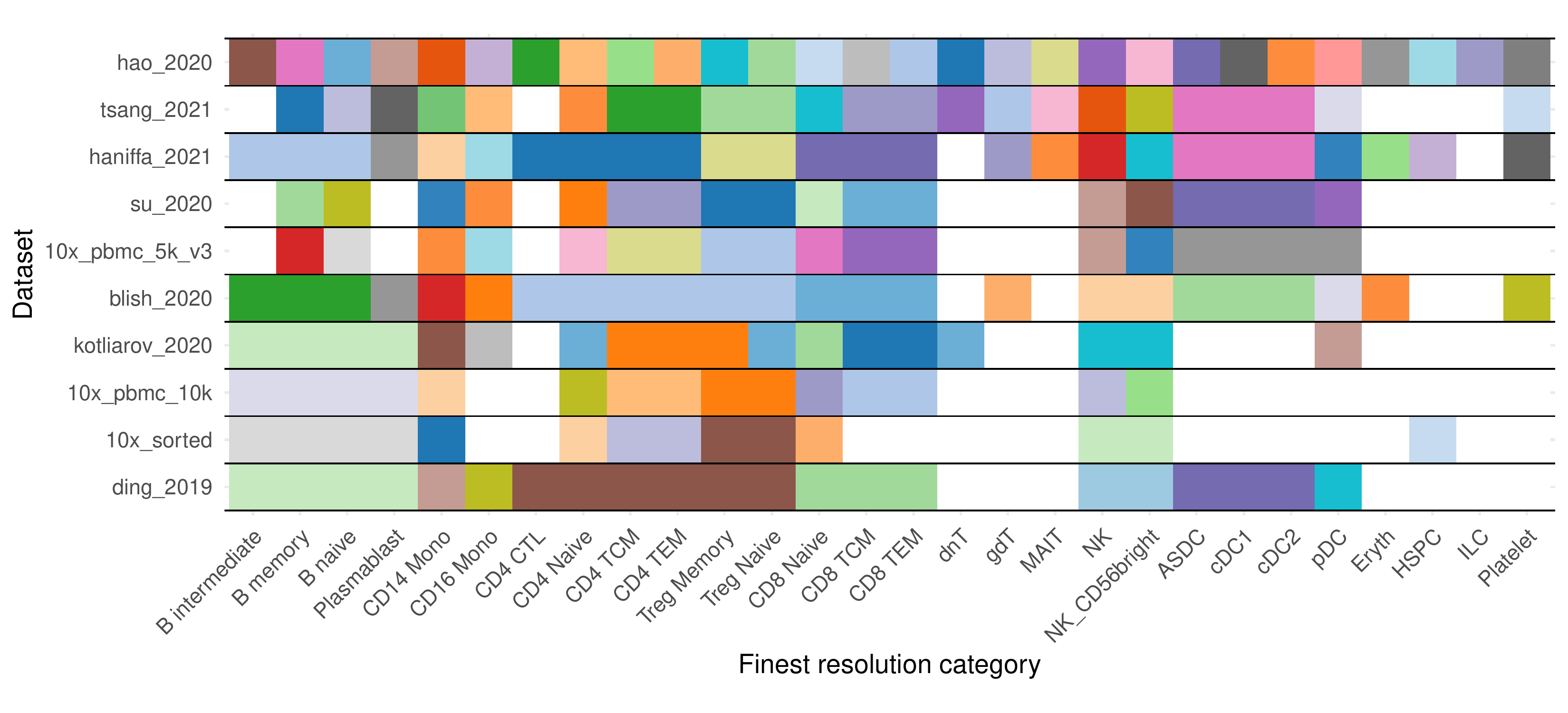}
\end{center}
\vspace{-25pt}
\caption{Graphical representation of the relationship between observed (annotated) labels and the finest resolution categories for each of the ten datasets from our integrative analysis in Section \ref{sec:data_analysis}. Within each row, when a color spans multiple finest resolution categories (columns), this indicates cells of these fine resolution categories were ``binned'' into a broader annotation label (coarse category) represented by the color. For example, in the \texttt{ding_2019} dataset (bottom row), each cell was annotated with one of eight distinct labels. One of these labels was ``B cell'' (represented by a pastel green color), and cells which could be described in detail as one of either ``B intermediate'', ``B memory,'' ``B naive,'' or ``Plasmablast'' are binned into the coarser ``B cell'' label. White spaces denote finest resolution categories which were not represented by the observed labels in a particular dataset.} \label{fig:data_cell_types}
\end{figure}

Our motivation for this work was to build a new and generalizable model for high-resolution cell type annotations for peripheral blood mononuclear cell (PBMC) samples by combining many publicly available datasets. We collected and processed a total of ten datasets sequenced using 10x Genomics technology, each with raw gene expression counts and curated cell type annotations available for each cell. We chose to work with PBMC data due to the complexity and hierarchy of immune cell types, as well as the common application of single-cell sequencing of PBMCs in clinical studies \citep{su2020multi, haniffastephenson2021single, blishwilk2020single}. Each of the ten datasets have labels at different resolutions, and although labels do not follow a tree-structure across datasets, they are amenable to binning. The number of distinct labels in each dataset, as well as references for the dataset, are shown in Table \ref{table:datasets}. The specific labels for each dataset are in Table 3 of the Supplementary Material. We display the relationships between labels represented in each of these datasets in Figure \ref{fig:data_cell_types} as graphical representations of ``binning functions,'' which are further described in Section \ref{subsec:binning_function}. The datasets we use are available through the R package \texttt{AnnotatedPBMC} at \url{https://github.com/keshav-motwani/AnnotatedPBMC/}, where we also provide an interface to our fitted model for predicting cell types from new single-cell gene expression data.

\section{Model}
%\subsection{Preliminaries}
Suppose we observe $K \geq 1$ datasets with single-cell gene expression profiles and cell types manually annotated. Let $\mathcal{C}_k$ denote the set of labels used to annotate the $k$th dataset for $k \in [K] = \{1,\dots, K\}$ and let $\mathcal{C}$ denote the set of labels at the desired finest resolution across all datasets. 
%The set $\mathcal{C}$ is user-specified, but cannot include any labels which do not belong to $\mathcal{C}_k$ for at least one $k \in [K]$ \textcolor{red}{This statement is confusing if $C_k$ and $C$ are numeric. the workaround is to say $\forall l \in C, \exists k, j \in C_k: g_k(j) = \{l\} $ but $g_k$ hasn't been defined yet. Keshav (again): I added a little blurb two paragraphs down at the end to hopefully resolve this}. 
Let $Y_{(k)i}$ and $\tilde{Y}_{(k)i}$ be the random variables corresponding to the annotated cell type and true (according to the finest resolution label set) cell type of the $i$th cell in the $k$th dataset for $i \in [n_k] = \{1, \dots, n_k\}$, $k \in [K]$, with supports $\mathcal{C}_k$ and $\mathcal{C}$, respectively. For the remainder, let $|\mathcal{A}|$ denote the cardinality of a set $\mathcal{A}$. Let $\mbX_{(k)} = (\mbx_{(k)1}, \dots, \mbx_{(k)n_k})^\top \in \mathbb{R}^{n_k \times p}$ be the observed gene expression matrix, and $(y_{(k)1}, \dots, y_{(k)n_k})^\top \in \mathcal{C}_k^{n_k}$ be a vector of cell type annotations for the $k$th dataset where $y_{(k)i}$ is the observed realization of the random variable $Y_{(k)i}$. Similarly, let $\tilde{\mbX}_{(k)} = (\tilde{\mbx}_{(k)1}, \dots, \tilde{\mbx}_{(k)n_k})^\top \in \mathbb{R}^{n_k \times p}$ for $k \in [K]$ be the unobservable gene expression matrix which is free of batch effects. Our goal is to estimate probabilities $P(\tilde{Y} = l | \mbx)$ for $l \in \mathcal{C}$ and any $\mbx \in \mathbb{R}^p$.

% and $\tilde{y}_{(k)} = (\tilde{y}_{(k)1}, \dots, \tilde{y}_{(k)n_k}) \in \mathcal{C}^{n_k}$ be a vector of unobservable finest resolution cell type labels where $\tilde{y}_{(k)i}$ is a realization of the random variable $\tilde{Y}_{(k)i}$.

\subsection{Binned categorical responses}\label{subsec:binning_function}
As described earlier, each dataset may have a different degree of resolution in their cell type annotations. Again taking an informal example, we may have two datasets with observed cell type labels in $\mathcal{C}_1 = \{A, B_1, B_2, B_3\}$ for the first dataset and $\mathcal{C}_2 = \{A_1, A_2, B\}$ for the second dataset, with $\mathcal{C} = \{A_1, A_2, B_1, B_2, B_3\}$ being the set of finest categories at which resolution we want to make predictions. %Abusing notation slightly, we use symbolic versions of the $\mathcal{C}$ (which are numeric) for the purpose of illustration.  
We refer to the labels $A$ and $B$ as ``coarse labels'' since groups of cells with these labels can each be partitioned into finer, more detailed categories (cells with label $A$ can be further divided into categories $A_1$ or $A_2$ and cells with label $B$ can be further divided into categories $B_1$, $B_2$, or $B_3$), and refer to each of $\{A_1, A_2, B_1, B_2, B_3\}$ as ``fine labels'' since they cannot be divided any further into more detailed categories. We refer to data observed at the level of a coarse label as a binned observation, because labels from finer categories are binned into one coarser label. For example, cells that are truly of cell type $A_1$ and $A_2$ are both binned into a label called $A$ in the first dataset. We will now make these ideas and definitions more formal by setting up some additional notation. 

Define the user-specified binning function $f_k: \mathcal{C} \to \mathcal{C}_k$ which maps a finest resolution category to the label used to describe that category in the $k$th dataset. For example, $f_1(A_1) = A$ for dataset 1 above. This function bins fine categories together into the possibly coarser resolution labels which are used in annotating the data, hence the name. Also, define the ``unbinning'' function $g_k = f^{-1}_k$ (inverse image) where $g_k(j) = f^{-1}_k(j) = \{l \in \mathcal{C} : f_k(l) = j \}$ for $j \in \mathcal{C}_k$. This provides the set of fine categories to which a cell labeled at a coarser resolution category may be further categorized as. For fine categories that are truly not represented in a given dataset, $f_k$ can map from these categories to another label (named ``unobserved'' for example). While $\mathcal{C}$ and the binning functions $f_k$ are user-specified, they must satisfy the condition that for all $l \in \mathcal{C}$, there must exist $k \in [K]$ and $j \in \mathcal{C}_k$ such that $f_k^{-1}(j) = g_k(j) = \{l\}$ with $\sum_{i = 1}^{n_k} \mathbb{1}(y_{(k)i} = j) \geq 1$. In other words, each of the finest resolution categories must actually be observed at least once in at least one of the $K$ training datasets. 
%\textcolor{red}{Do the last two sentences make sense? Moved this idea from above}

% Define the ``unbinning" function $g_k: \mathcal{C}_k \to \mathcal{P}(\mathcal{C})$,- where $\mathcal{P}(\cdot)$ denotes the power set of its argument, which maps a coarse category from the $k$th dataset to the set of fine subcategories of it. For example, in the toy example described in the previous paragraph, $g_1$ would map $A$ to $\{A_1, A_2\}$ and $B_l$ to $B_l$ for $l = 1, 2, 3.$
% We will require only that $g_k(j) \cap g_k(j') = \emptyset$ for all sets $j, j' \in \mathcal{C}_k$ and $\bigcup_{j \in \mathcal{C}_k} g_k(j) = \mathcal{C}$ $(k = 1,\dots, K)$, i.e. that each fine category belongs to only one coarse category and each of the fine categories are represented by one coarse category in each dataset. If some fine categories are not represented by a coarse category in a particular dataset, we can simply add a category ``other'' to $\mathcal{C}_k$ to create $\tilde{\mathcal{C}}_k = \mathcal{C}_k \cup \left\{ \text{``other''} \right\}$ and define $\tilde{g}_k: \tilde{\mathcal{C}}_k \to \mathcal{P}(\mathcal{C})$ such that
% $$
% \tilde{g}_k(j) = 
% \begin{cases}
% g_k(j), & j \in \mathcal{C}_k,\\
% \mathcal{C} \setminus \bigcup_{j' \in \mathcal{C}_k} g_k(j'), & j = \text{``other''}
% \end{cases}
% $$
% and use $\tilde{\mathcal{C}}_k$ and $\tilde{g}_k$ in place of $\mathcal{C}_k$ and $g_k$. 

Using this notation, we can now formally define $j \in \mathcal{C}_k$ to be a ``coarse label'' if $|g_k(j)| > 1$ (i.e., the label can be broken up into multiple finer resolution categories) and a ``fine label'' if $|g_k(j)| = 1$ (i.e., the label cannot be further partitioned). We also now define the relationship between $Y_{(k)i}$ and $\tilde{Y}_{(k)i}$ through the following equivalence of events
% $$ \{Y_{(k)i} = j\} = \bigcup_{l \in g_k(j)} \{ \tilde{Y}_{(k)i} = l\}, ~~~ j \in \mathcal{C}_k.$$
$$ \{Y_{(k)i} = j\} = \bigcup_{l \in g_k(j)} \{ \tilde{Y}_{(k)i} = l\}, ~~~ j \in \mathcal{C}_k.$$
That is, a cell can be categorized within one of the finest resolution categories in the bin corresponding to the observed label, with the correspondence defined by $g_k$.
%This represents the idea that the true, finer resolution cell type is one of the subcategories of the labeled cell type. 
We thus have that 
\begin{equation} \label{eq:summed_probabilities}
P(Y_{(k)i} = j \mid \mbx_{(k)i}) = \hspace{-2pt}\sum_{l \in g_k(j)} \hspace{-2pt}P(\tilde{Y}_{(k)i} = l\mid \mbx_{(k)i}) , ~~~ j \in \mathcal{C}_k
\end{equation}
since the events $\{ \tilde{Y}_{(k)i} = l: l \in g_k(j)\}$ are mutually exclusive as a cell can only be of one cell type.

\subsection{Binned multinomial regression model}
As mentioned, we are interested in modeling cell type probabilities as a function of gene expression. For now, we consider a model using unobserved gene expression $\tilde{\mbx}_{(k)i}$, which is free of batch effects, and will extend this in the next section to the observed gene expression. Without loss of generality, we encode the sets of labels numerically so that $\mathcal{C} = \{1, \dots, |\mathcal{C}|\}$ and $\mathcal{C}_k = \{1, \dots, |\mathcal{C}_k|\}$ for $k \in [K].$
We assume that each $\tilde{Y}_{(k)i}$ follows a categorical distribution (i.e., multinomial based on a single trial)
$$\tilde{Y}_{(k)i} \sim {\rm Categorical}\{\pi_1^*(\tilde{\mbx}_{(k)i}), \dots, \pi_{|\mathcal{C}|}^*(\tilde{\mbx}_{(k)i})\}.$$
In addition, we assume that the probability functions $\pi_l^*$ adhere to the standard multinomial logistic regression link so that
\begin{equation} \label{eq:multinomial_probability}
\pi_l^*(\tilde{\mbx}_{(k)i}) = \frac{{\rm exp}(\mbalpha_l^* + \tilde{\mbx}_{(k)i}^\top \mbbeta_{l}^*)}{\sum_{v \in \mathcal{C}} {\rm exp}(\mbalpha_{v}^* +\tilde{\mbx}_{(k)i}^\top \mbbeta_{v}^*)}, ~~~  l \in \mathcal{C}, ~~~ k \in [K],
\end{equation}
% $$ P(\tilde{Y}_{(k)i} = l \mid \tilde{x}_{(k)i}) = \frac{{\rm exp}(\alpha_l + \beta_{l}'\tilde{x}_{(k)i})}{\sum_{v \in \mathcal{C}} {\rm exp}(\alpha_{v} +\beta_{v}'\tilde{x}_{(k)i})},~~~ l \in \mathcal{C},$$
where $\mbalpha^* = (\mbalpha^*_1, \dots, \mbalpha^*_{|\mathcal{C}|})^\top \in \mathbb{R}^{|\mathcal{C}|}$ is an unknown vector of intercepts and $\mbbeta^* = (\mbbeta^*_1, \dots, \mbbeta^*_{|\mathcal{C}|}) \in \mathbb{R}^{p \times |\mathcal{C}|}$ is an unknown matrix of regression coefficients. Applying exactly the logic from \eqref{eq:summed_probabilities}, it follows that
$$ P(Y_{(k)i} = j \mid \tilde{\mbx}_{(k)i}) = \sum_{l \in g_k(j)} \pi_l^*(\tilde{\mbx}_{(k)i}) = \frac{\sum_{l \in g_k(j)}{\rm exp}(\mbalpha_l^* + \tilde{\mbx}_{(k)i}^\top \mbbeta_{l}^*)}{\sum_{v \in \mathcal{C}} {\rm exp}(\mbalpha_{v}^* +\tilde{\mbx}_{(k)i}^\top \mbbeta_{v}^*)}, ~~~ j \in \mathcal{C}_k, ~~~ k \in [K].$$ 
Thus, our focus is the development of a method for estimating $\mbalpha^*$ and $\mbbeta^*$. However, we first extend the model to account for potential batch effects in the observed gene expression. 

\subsection{Adjustment for batch effects}\label{subsec:batch_effects}
The gene expression $\mbx_{(k)i}$ can be assumed to be ``noisy'' in the sense that they may be measured with some batch effects specific to each of the $K$ datasets. For example, it may be reasonable to assume that $\mbx_{(k)i} = \tilde{\mbx}_{(k)i} + \mbu_{(k)i}$ where $\tilde{\mbx}_{(k)i}$ is the the unobserveable gene expression and $\mbu_{(k)i}$ is some noise. This additive assumption of batch effects is consistent with the existing literature on data integration for normalized gene expression data in single-cell datasets, which provide methods for estimating the $\mbu_{(k)i}$ \citep{haghverdi2018batch,hao2020integrated}. However, estimating the per-gene batch effect is not necessary for classification: we need only estimate a linear combination of this batch effect, as we now describe. 

We can write the linear predictor for the $i$th cell of the $k$th dataset as
$\mbalpha^* + \tilde{\mbx}_{(k)i}^\top \mbbeta^*_l = \mbalpha^* + \mbx_{(k)i}^\top\mbbeta^*  - \mbu_{(k)i}^\top \mbbeta^*.$
Because the $\mbu_{(k)i}$ are not observable, we assume that there are some common sources of batch variation which are related to some cell-specific covariates $\mbz_{(k)i} \in \mathbb{R}^r$, and that $\mbu_{(k)i}$ is some linear combination of these cell specific covariates $\mbu_{(k)i} = \mbz_{(k)i}^\top \mbphi_{(k)}^*$ for $i \in [n_k]$, $k \in [K]$, and coefficients $\mbphi_{(k)}^* \in \mathbb{R}^{r \times p}$. It follows that the linear predictor for the $i$th cell in the $k$th dataset is
$\mbalpha^* + \tilde{\mbx}_{(k)i}^\top \mbbeta^* = \mbalpha^*  + \mbx_{(k)i}^\top \mbbeta^* -  \mbz_{(k)i}^\top \mbphi^*_{(k)} \mbbeta^*$
where $\mbalpha^*$, $\mbbeta^*$, and the $\mbphi^*_{(k)}$ are unknown. Letting $\mbgamma^*_{(k)} = -\mbphi^*_{(k)}\mbbeta^*$ (since both are unknown), we can see that 
$\mbalpha^* + \tilde{\mbx}_{(k)i}^\top\mbbeta^*  = \mbalpha^* + \mbx_{(k)i}^\top \mbbeta^* + \mbz_{(k)i}^\top \mbgamma_{(k)}^* .$
Thus, we can write
\begin{equation} \label{eq:probabilities}
P(Y_{(k)i} = j \mid \mbx_{(k)i}, \mbz_{(k)i}) = \sum_{l \in g_k(j)}\frac{{\rm exp}(\mbalpha^*_l + \mbx_{(k)i}^\top \mbbeta_{l}^* +  \mbz_{(k)i}^\top\mbgamma^*_{(k)l})}{\sum_{v \in \mathcal{C}} {\rm exp}(\mbalpha^*_{v} + \mbx_{(k)i}^\top \mbbeta_{v}^* + \mbz_{(k)i}^\top \mbgamma^*_{(k)v})}, ~~~ j \in \mathcal{C}_k, ~~~ k \in [K]
\end{equation}
In the simplest case, $\mbz_{(k)i} = 1$ (i.e., provides an intercept adjustment), which implies a batch-specific shift in expression that is constant for all cells in the batch. Alternatively, $\mbz_{(k)i}$ can also contain the principal components of $(\mbX_{(1)}^\top, \dots, \mbX_{(K)}^\top)^\top$ to capture interactions of batch with other directions of variation in the data. It is worth emphasizing that here, we have both batch specific coefficients to estimate, $\mbgamma^*_{(k)}$ for $k \in [K]$, and coefficients shared across batches, ($\mbalpha^*, \mbbeta^*).$
With this, our goal will be to estimate $\mbalpha^*$, $\mbbeta^*$, and $\mbgamma^*_{(k)}$ via penalized maximum likelihood based on the observed predictors $\mbx_{(k)i}$ for $i \in [n_k]$ and $k \in [K].$

\section{Methodology}
\subsection{Penalized maximum likelihood estimator}
From the probability functions described in Section \ref{subsec:batch_effects}, we see that the log-likelihood contribution for the $i$th cell in the $k$th dataset can be expressed 
$$
l_{(k)i}(\mbalpha, \mbbeta, \mbgamma_{(k)}) =  \sum_{j \in \mathcal{C}_k} \mathbb{1}(y_{(k)i} = j) \log \left( \sum_{l \in g_k(j)}\frac{{\rm exp}(\mbalpha_l + \mbx_{(k)i}^\top \mbbeta_{l}  + \mbz_{(k)i}^\top \mbgamma_{(k)l})}{\sum_{v \in \mathcal{C}} {\rm exp}(\mbalpha_{v} + \mbx_{(k)i}^\top \mbbeta_{v} + \mbz_{(k)i}^\top \mbgamma_{(k)v})} \right)
$$
for $i \in [n_k]$ and $k\in[K]$, where $\mathbb{1}$ denotes the indicator function. 
We can therefore define the (scaled by $1/N$) negative log-likelihood as
$$\mathcal{L}(\mbalpha, \mbbeta, \mbgamma) = - \frac{1}{N} \sum_{k = 1}^{K} \sum_{i = 1}^{n_k} l_{(k)i}(\mbalpha, \mbbeta, \mbgamma_{(k)}),$$
where $N = \sum_{k=1}^K n_k$ is the total sample size and $\mbgamma = (\mbgamma_{(1)}, \dots, \mbgamma_{(K)}) \in \mathbb{R}^{r \times |\mathcal{C}|} \times \cdots \times \mathbb{R}^{r \times |\mathcal{C}|}$. We thus estimate $\mbalpha^*$ and $\mbbeta^*$, which are the shared across datasets, and $\mbgamma^*_{(k)} \in \mathbb{R}^{r \times |\mathcal{C}|}$ for datasets $k \in [K]$ jointly using penalized maximum likelihood. For ease of display, let $\mathcal{T} = \mathbb{R}^{|\mathcal{C}|} \times \mathbb{R}^{p \times |\mathcal{C}|} \times \mathbb{R}^{r \times |\mathcal{C}|} \times \cdots \times \mathbb{R}^{r \times |\mathcal{C}|}$ be the space of the unknown parameters $(\mbalpha^*, \mbbeta^*, \mbgamma^*).$ Formally, the estimator of $(\mbalpha^*, \mbbeta^*, \mbgamma^*)$ we propose is
\begin{equation}\label{eq:estimator} \argmin_{(\mbalpha, \mbbeta, \mbgamma) \in\mathcal{T} } \left\{\mathcal{L}(\mbalpha, \mbbeta, \mbgamma) + \lambda \sum_{j=1}^p \|\mbbeta_{j,:}\|_2 \hspace{3pt} + \hspace{3pt}\frac{\rho}{2}\sum_{k=1}^K \|\mbgamma_{(k)}\|_F^2\right\},
\end{equation}
where $\mbbeta_{j,:} \in \mathbb{R}^{|\mathcal{C}|}$ denotes the $j$th row of $\mbbeta$ for $\in[p] = \{1, \dots, p\}$, $\|\cdot\|_2$ denotes the Euclidean norm of a vector, $\|\cdot\|_F$ denotes the Frobenius norm of a matrix, and $(\lambda, \rho) \in (0, \infty) \times (0, \infty)$ are user-specified tuning parameters. We now motivate the choice of penalties based on our application.

Manual single-cell annotation is often performed through the identification of upregulated genes within clusters of cells \citep{amezquita2020orchestrating}. For example, to label a cluster of cells as type CD4+ naive, an annotater often identifies a number of particular genes that are overexpressed in that cluster relative to the rest of the cells \citep{wolf2018scanpy, hao2020integrated}. This implies that a relatively small number of genes are necessary to characterize the relationship between cell type probabilities and gene expression. For this reason we use the group lasso type penalty on the rows of the optimization variable $\mbbeta$ \citep{yuan2006model,obozinski2011support,simon2013sparse}. For large values of $\lambda$, this penalty will encourage estimates of $\mbbeta^*$ which will have rows either entirely equal to zero or entirely nonzero. If the $j$-th row of $\mbbeta^*$ is zero, the $j$-th gene is irrelevant for discriminating between cell types. The $L_{1}$(vector)-norm penalty (i.e., the lasso penalty), in contrast, would not lead to easily interpreted variable selection since a zero in a particular entry of $\mbbeta^*$ does not alone imply anything about whether the corresponding predictor affects the probabilities.

Regarding the ridge penalty on the $\mbgamma_{(k)}$: because the  $\mbgamma_{(k)}$ are specific to each of the training sets, we do not have corresponding coefficients for a test data point from a new (i.e., unobserved for training) dataset. Additionally, we expect that the batch effect does not contain information relevant to cell type classification. Therefore, we intuitively want $\mbgamma_{(k)}$ to be close to the origin, so that on a test data point, we can simply use our estimates $\hat{\mbalpha}$ and $\hat{\mbbeta}$ from \eqref{eq:estimator} to estimate probabilities with 
$$ \hat{P}(\tilde{Y} = l \mid \mbx) = \frac{{\rm exp}(\hat{\mbalpha}_l + \mbx^\top \hat{\mbbeta}_{l})}{\sum_{v \in \mathcal{C}} {\rm exp}(\hat{\mbalpha}_{v} + \mbx^\top \hat{\mbbeta}_{v})}, ~~~ l \in \mathcal{C},$$
as if $\tilde{\mbx} = \mbx$.
To encourage estimates of the $\mbgamma^*_{(k)}$ to be small, we add a penalty of the squared Frobenius norm of each $\mbgamma_{(k)}$. Additional intuition may be gleaned by considering the Bayesian interpretation of ridge regression wherein the coefficients are assumed to follow a mean zero normal distribution.  %\citet{parker2012practical}, for example, found that batch effects have little effect on prediction when batch is uncorrelated with the response. In our setting, it is reasonable to assume that batch effects are relatively uncorrelated with cell types.  

Importantly, the coefficients we intend to estimate are not, in general, identifiable. This is because with $\mathbf{1}_{|\mathcal{C}|} = (1, \dots, 1)^\top \in \mathbb{R}^{|\mathcal{C}|},$ for any $(\mbalpha, \mbbeta, \mbgamma)$,  $\mathcal{L}(\mbalpha, \mbbeta, \mbgamma_{(1)}, \dots, \mbgamma_{(K)}) = \mathcal{L}(\mbalpha - a\cdot \mathbf{1}_{\mathcal{|C|}}^\top, \mbbeta- \boldsymbol{b} \mathbf{1}_{\mathcal{|C|}}^\top, \mbgamma_{(1)} - \boldsymbol{d}_1 \mathbf{1}_{\mathcal{|C|}}^\top, \dots, \mbgamma_{(K)}- \boldsymbol{d}_K \mathbf{1}_{\mathcal{|C|}}^\top)$
for any $a \in \mathbb{R}$, $\boldsymbol{b} \in \mathbb{R}^p$, and $\boldsymbol{d}_k \in \mathbb{R}^r$ for $k \in [K]$. However, if we impose the ``sum-to-zero" condition that $\mbalpha^\top\mathbf{1}_{|\mathcal{C}|} = \mbbeta_{1,:}^\top \mathbf{1}_{|\mathcal{C}|} = \cdots = \mbbeta_{p,:}^\top \mathbf{1}_{|\mathcal{C}|} = 0,$ and similarly for the rows of the $\mbgamma_{(k)}$, then this issue may be resolved. It is perhaps surprising that the $\mbgamma_{(k)} \in \mathbb{R}^{r \times |\mathcal{C}|}$ could be identifiable since $\mathcal{C}_k$ may be distinct from $\mathcal{C}$, but one can see that replacing $\mbgamma_{(k)}$ with $\mbgamma_{(k)}'$ will, in general, lead to distinct probabilities \eqref{eq:probabilities} unless $\mbgamma_{(k)}' = \mbgamma_{(k)} - \boldsymbol{d}_k \mathbf{1}_{|\mathcal{C}|}^\top$. In the Supplementary Material, we discuss the (exceptionally rare) situations where this is not true.  Fortunately, both our penalties naturally enforce the sum-to-zero constraints on $\mbbeta$ and the $\mbgamma_{(k)}$. For example, see the Supplementary Material of \citet{molstad2021likelihood} for a proof of this fact. 

% \subsection{Identifiability}
% It is natural to ask when the parameters $(\mbalpha, \mbbeta, \mbgamma)$ are identifiable. Of course, without any constraints, it is easy to see that they are not. For example, suppose we excluded batch effects and observed responses at the leve of the finest categories in each dataset. In this case, it is easy to see that if we replaced $\mbalpha$ with $\mbalpha_c = \mbalpha - c 1_{|\mathcal{C}|}$ and replaced $\mbbeta$ with $\mbbeta_d = \mbbeta - \mbd 1_{|\mathcal{C}|}^\top$ for any $c \in \mathbb{R}$ and any $\mbd \in \mathbb{R}^p$, $(\mbalpha, \mbbeta)$ and $(\mbalpha_c, \mbbeta_d)$ would yield the same probabilities. A standard rememdy is to require that the components of $\mbalpha$ and the components of each row of $\mbbeta$ sum to zero.  In fact, we automatically enforce a sum-to-zero condition in our setting and, under some mild conditions on the granularity of the responses. 
% \begin{remark}
% Suppose $(\hat\mbalpha, \hat\mbbeta, \hat\mbgamma)$ are as defined in \eqref{eq:estimator} for $\lambda > 0$ and $\gamma > 0$
% \textcolor{blue}{Keshav, here we basically want to argue that under the sum of zero constraints, the $(\hat\mbalpha, \hat\mbbeta, \hat\mbgamma)$ are identifiable. }
% \end{remark}

\subsection{Related methods}
The approach proposed here is closely related to a growing literature on methods for integrative analyses. We discuss this literature from two perspectives: that of statistical methodology and that of the analysis of multiple single-cell datasets jointly.

From a methodological perspective, there is a growing interest in developing methods for jointly analyzing datasets from heterogeneous sources. Most often, these methods assume distinct data generating models for each source and aim improve efficiency by exploiting similarities across sources \citep{zhao2015integrative,huang2017promoting,ventz2021integration,molstad2021dimension}. For example, \citet{huang2017promoting} assumed a similar sparsity pattern for regression coefficients corresponding to separate populations. Similarly, \citet{molstad2021dimension} assumed a shared low-dimensional linear combination of predictors explained the outcome in all sources. The focus of our work is different: the sources from which the data were collected are assumed to differ only in their response category label resolution (and, to a lesser degree, may measure predictors with batch effects). Thus, these approaches are, generally speaking, not directly applicable to our setting. 

In the context of single-cell data analysis, integrative analyses often focus on the ``alignment'' of expression datasets in an attempt to remove batch effects for the purposes of clustering and visualization \citep{haghverdi2018batch,hie2019efficient,korsunsky2019fast,hao2020integrated}. As mentioned in the previous section, explicit estimation and removal of batch effects is not necessary for the goal of cell type prediction. In fact, \citet{ma2021evaluation} found that removing batch effects through alignment-based methods actually decreased downstream cell type prediction accuracy.
Our inclusion of batch specific  effects in \eqref{eq:probabilities} can, loosely speaking, be thought of as performing alignment specifically tailored to prediction (assuming the $\mbz_{(k)i}$ are chosen appropriately).

\section{Computation}
%\subsection{Overview}
In order to compute our proposed estimator, we must address that the group lasso penalty is nondifferentiable at zero and that the overall negative log-likelihood $\mathcal{L}$ is nonconvex in general. In brief, we employ a blockwise proximal gradient descent scheme \citep{xu2017globally} to overcome these challenges. Specifically, we obtain a new iterate by minimizing a penalized quadratic approximation to $\mathcal{L}$ at the current iterate, which will ensure -- by the majorize-minimize principle \citep{lange2016mm} -- a monotonically decreasing objective function value. Our approximations are chosen so as to admit simple, closed form updates for each block. In the remainder of this section, we motivate and derive each block update and summarize our algorithm. %In the Supplementary Material, we provide some important details about our implementation including a discussion on how we determine candidate tuning parameter sets. 
Code implementing the algorithm described here is available for download at \url{https://github.com/keshav-motwani/IBMR/}.

%\subsection{Blockwise proximal gradient descent algorithm}
Let $\mathcal{F}_{\lambda, \rho}$ denote the objective function from \eqref{eq:estimator}. By construction, $\mathcal{F}_{0, 0}$ denotes the negative log-likelihood $\mathcal{L}.$  To describe our iterative procedure, we focus on the update for $\mbbeta$, but as we will show, this approach also applies to $\mbalpha$ and the $\mbgamma_{(k)}$ with minor modification. First, notice that given $t$-th iterates of $\mbalpha, \mbbeta$ and $\mbgamma$, $(\mbalpha^t,\mbbeta^t, \mbgamma^t)$, by the Lipschitz continuity of the gradient of $\mathcal{L}$ with respect to $\mbbeta$, we know that for any step size $s_\beta$ such that $0 < s_\beta <  N/\{\sqrt{|\mathcal{C}|}\sum_{k = 1}^{K} \| \mbX_{(k)} \|_F^2\}$,
\begin{align} \label{eq:lipschitzquadratic}
\mathcal{F}_{0, 0}(\mbalpha^{t}, \mbbeta, \mbgamma^{t}) & \leq \mathcal{F}_{0, 0}(\mbalpha^{t}, \mbbeta^{t}, \mbgamma^{t})  + {\rm tr}\left\{\nabla_\beta \mathcal{F}_{0, 0}(\mbalpha^{t}, \mbbeta^t, \mbgamma^t)^\top (\mbbeta - \mbbeta^t)\right\} + \frac{1}{2s_\beta}\|\mbbeta - \mbbeta^t\|_F^2
\end{align} 
for all $\mbbeta \in \mathbb{R}^{p \times |\mathcal{C}|}$,
where $\nabla_\beta \mathcal{F}_{0,0}(\mbalpha^t, \cdot, \mbgamma^t)$ denotes the gradient of $\mbbeta \mapsto \mathcal{F}_{0, 0}(\mbalpha^t, \mbbeta, \mbgamma^t)$. 
Letting $\mathcal{M}(\mbbeta \mid \mbbeta^t)$ denote the right-hand side of the above inequality, we can see that 
$$ \mathcal{F}_{\lambda, \rho}(\mbalpha^{t}, \mbbeta, \mbgamma^{t}) \leq \mathcal{M}(\mbbeta \mid \mbbeta^{t}) +  \lambda \sum_{j=1}^p\|\mbbeta_{j,:}\|_2 \hspace{2pt} + \hspace{2pt}\frac{\rho}{2}\sum_{k=1}^K \|\mbgamma_{(k)}^t\|_F^2,$$
for all $\mbbeta \in \mathbb{R}^{p \times |\mathcal{C}|}$ with equality when $\mbbeta = \mbbeta^t$. If we thus define $\mbbeta^{t+1}$ as the argument minimizing $\mathcal{M}(\mbbeta\mid  \mbbeta^t) + \lambda \sum_{j=1}^p\|\mbbeta_{j,:}\|_2$, we are ensured that $\mathcal{F}_{\lambda,\gamma}(\mbalpha^{t}, \mbbeta^{t+1}, \mbgamma^{t}) \leq \mathcal{F}_{\lambda,\gamma}(\mbalpha^{t}, \mbbeta^{t}, \mbgamma^{t}).$
Hence, defining $\mbbeta^{t+1}$ in this way, we have
\begin{align*}
\mbbeta^{t+1} & = \argmin_{\mbbeta \in \mathbb{R}^{|\mathcal{C}|}} \left\{\mathcal{M}(\mbbeta\mid \mbbeta^t) + \lambda \sum_{j=1}^p\|\mbbeta_{j,:}\|_2\right\}  = \argmin_{\mbbeta \in \mathbb{R}^{|\mathcal{C}|}} \left\{ \frac{1}{2}\|\mbbeta - \mbnu^t(s_\beta)\|_F^2 + s_\beta\lambda \sum_{j=1}^p\|\mbbeta_{j,:}\|_2\right\},
\end{align*}
where $\mbnu^t(s_\beta) = \mbbeta^{t} - s_\beta \nabla_\beta \mathcal{F}_{0,0}(\mbalpha^{t}, \mbbeta^{t}, \mbgamma^{t})$. The second equality above implies that $\mbbeta^{t+1}$ is simply the proximal operator \citep{parikh2014proximal,polson2015proximal} of the $\|\cdot\|_{1,2}$-norm  (sum of Euclidean norms of the rows of its matrix-valued argument) at $\mbnu^t(s_\beta)$. Some straightforward derivations (e.g., see \citet{simon2013sparse}) reveal that the $j$th row of $\mbbeta^{t+1}$, $\mbbeta^{t+1}_{j,:},$ can thus be obtained in closed form 
$$ \mbbeta^{t+1}_{j,:} = \max\left(1 - \frac{s_\beta \lambda}{\|\mbnu^t(s_\beta)_{j,:}\|_2}, 0 \right)\mbnu^t(s_\beta)_{j,:}, ~~~j \in [p].$$
We apply analogous arguments to update both $\mbgamma$ with $(\mbalpha^{t}, \mbbeta^{t+1})$ fixed and $\mbalpha$ with $(\mbbeta^{t+1}, \mbgamma^{t+1})$ fixed. For $\mbalpha$, yields a standard gradient descent update, whereas for the $\mbgamma_{(k)}$, each can be updated in parallel. Specifically, by the same motivation as in the update for $\mbbeta$, we define 
\begin{align*}
\mbgamma_{(k)}^{t+1} &= \argmin_{\mbgamma_{(k)} \in \mathbb{R}^{r \times |\mathcal{C}|}} \left\{ \frac{1}{2}\|\mbgamma_{(k)} - \mbgamma_{(k)}^{t} + s_{\gamma_{(k)}} \nabla_{\gamma_{(k)}} \mathcal{F}_{0,0}(\mbalpha^{t}, \mbbeta^{t+1}, \mbgamma^{t})\|_F^2 + \frac{s_{\gamma_{(k)}}\rho}{2} \|\mbgamma_{(k)}\|_F^2\right\}\\
& = \left(1 + s_{\gamma_{(k)}} \rho \right)^{-1}\left\{\mbgamma_{(k)}^{t+1} -  s_{\gamma_{(k)}} \nabla_{\gamma_{(k)}} \mathcal{F}_{0,0}(\mbalpha^{t}, \mbbeta^{t+1}, \mbgamma^{t})\right\}.
\end{align*}
With these updating expressions for $\mbbeta, \mbalpha,$ and $\mbgamma$ in hand, we formally state our iterative procedure for minimizing $\mathcal{F}_{\lambda, \rho}$ in Algorithm \ref{alg:1}. 
Applying an identical series of arguments as those to prove that $\mbbeta^{t+1}$ yields a decrement of the objective function, we have the following lemma regarding the sequence of iterates $\{(\mbbeta^{t}, \mbalpha^t, \mbgamma^t)\}_{t=0}^\infty$.
\begin{lemma} (Descent property)
As long as each step size $s_\beta >0, s_\alpha >0 , s_{\gamma_{(k)}} > 0$ is sufficiently small and fixed or chosen by backtracking line search (see the Supplementary Material), the sequence of iterates $\{(\mbbeta^{t}, \mbalpha^t, \mbgamma^t)\}_{t=0}^\infty$ is guaranteed to satisfy
$ \mathcal{F}_{\lambda, \rho}(\mbalpha^{t+1}, \mbbeta^{t+1}, \mbgamma^{t+1}) \leq \mathcal{F}_{\lambda, \rho}(\mbalpha^{t}, \mbbeta^{t}, \mbgamma^{t}),$
for $t = 1, 2, 3, \dots,$ i.e., Algorithm \ref{alg:1} has the descent property. 
\end{lemma}

\begin{algorithm}[t] \caption{Blockwise proximal gradient descent algorithm for minimizing $\mathcal{F}_{\lambda, \rho}$}
    \label{alg:1}
    Initialize $\mbbeta^{0} \in \mathbb{R}^{p \times |\mathcal{C}|}$, $\mbalpha^{0} \in \mathbb{R}^{|\mathcal{C}|}$, and $\mbgamma^{0}_{(k)} \in \mathbb{R}^{r \times |\mathcal{C}|}$ for $k \in [K]$. Set $t = 0$.
\begin{enumerate}
\item Compute $\mbnu^t(s_\beta) = \mbbeta^{t} - s_\beta \nabla_\beta \mathcal{F}_{0,0}(\mbalpha^{t}, \mbbeta^{t}, \mbgamma^{t})$ 
\item For $j \in [p]$ in parallel, compute
$$\mbbeta^{t+1}_{j,:} =  \max\left(1 - \frac{s_\beta \lambda}{\|\mbnu^t(s_\beta)_{j,:}\|_2}, 0 \right)\mbnu^t(s_\beta)_{j,:} $$
with $s_\beta$ chosen by backtracking line search. 
\item For $k \in [K]$ in parallel, compute
$$\mbgamma_{(k)}^{t+1} = \left(1 + s_{\gamma_{(k)}} \rho \right)^{-1}\left\{\mbgamma_{(k)}^{t} -  s_{\gamma_{(k)}} \nabla_{\mbgamma_{(k)}} \mathcal{F}_{0,0}(\mbalpha^{t}, \mbbeta^{t+1}, \mbgamma^{t})\right\}$$
with the $s_{\gamma_{(k)}}$ chosen by backtracking line search. 
\item Compute $\mbalpha^{t+1} = \mbalpha^{t} - s_\alpha \nabla_{\alpha_{(k)}} \mathcal{F}_{0,0}(\mbalpha^{t}, \mbbeta^{t+1}, \mbgamma^{t+1})$ with $s_\alpha$ chosen by backtracking line search. 
\item If objective function value has not converged, set $t = t + 1$ and return to 1. 
\end{enumerate}
\end{algorithm}

In the Supplementary Material, we derive explicit forms of the partial derivatives needed in Algorithm \ref{alg:1}. Because they provide some insight, we discuss them here. 
For each $k \in [K]$, let $\tilde{\mbP}_{(k)}:\mathbb{R}^{|\mathcal{C}|} \times \mathbb{R}^{p \times |\mathcal{C}|} \times \mathbb{R}^{r \times |\mathcal{C}|} \times \cdots \mathbb{R}^{r \times |\mathcal{C}|} \to \mathbb{R}^{n \times |\mathcal{C}|}$ be a matrix-valued function which maps input parameters $(\mbalpha, \mbbeta, \mbgamma)$ to a matrix of (unconditional) probabilities.  Specifically, $\tilde{\mbP}_{(k)}(\mbalpha, \mbbeta, \mbgamma_{(k)})$ has $(i,l)$-th entry
\begin{equation} \label{eq:Pmatrix}
[\tilde{\mbP}_{(k)}(\mbalpha, \mbbeta, \mbgamma_{(k)})]_{i,l} = \frac{{\rm exp}(\mbalpha_l + \mbx_{(k)i}^\top \mbbeta_{l}  + \mbz_{(k)i}^\top \mbgamma_{(k)l})}{\sum_{v \in \mathcal{C}} {\rm exp}(\mbalpha_{v} + \mbx_{(k)i}^\top \mbbeta_{v} + \mbz_{(k)i}^\top \mbgamma_{(k)v})}, ~~~ l \in \mathcal{C},~~ i \in [n_k],~~ k \in [K].
\end{equation}
Similarly, let $\tilde{\mbC}_{(k)}:\mathbb{R}^{|\mathcal{C}|} \times \mathbb{R}^{p \times |\mathcal{C}|} \times \mathbb{R}^{r \times |\mathcal{C}|} \times \cdots \mathbb{R}^{r \times |\mathcal{C}|} \to \mathbb{R}^{n \times |\mathcal{C}|}$ be a matrix-valued function of conditional probabilities where
\begin{equation} \label{eq:Cmatrix}
[\tilde{\mbC}_{(k)}(\mbalpha, \mbbeta, \mbgamma_{(k)})]_{i,l} = \frac{\mathbb{1}\{l \in g_k(y_{(k)i})\}~{\rm exp}(\mbalpha_l + \mbx_{(k)i}^\top \mbbeta_{l}  + \mbz_{(k)i}^\top \mbgamma_{(k)l})}{ \sum_{v \in g_k(y_{(k)i})} {\rm exp}(\mbalpha_{v} + \mbx_{(k)i}^\top \mbbeta_{v} + \mbz_{(k)i}^\top \mbgamma_{(k)v})}, ~~~ l \in \mathcal{C}, ~~i \in [n_k],~~ k \in [K].
\end{equation}
Intuitively, $[\tilde{\mbP}_{(k)}(\mbalpha, \mbbeta, \mbgamma_{(k)})]_{i,l}$ is the estimated probability that cell $i$ from dataset $k$ is of type $l$. The conditional probability $[\tilde{\mbC}_{(k)}(\mbalpha, \mbbeta, \mbgamma_{(k)})]_{i,l}$ is the estimated probability that cell $i$ from dataset $k$ is of type $l \in \mathcal{C}$ given $y_{(k)i}$ is the observed (possibly coarse) label. Of course, if $g_k(y_{(k)i})$ is a singleton, then $[\tilde{\mbC}_{(k)}(\mbalpha, \mbbeta, \mbgamma_{(k)})]_{i,l} = \mathbb{1}\{l \in g_k(y_{(k)i})\}$. %\textcolor{red}{I changed this, see comment below in tex file.}
%$f_k(l)$ is a singleton (i.e., the fine category $l$ is measured on the $k$th dataset) \textcolor{red}{$f_k$ (the binning function) always returns singletons, I think what you meant here is the above If this isn't what you meant, let's discuss :)}, then this simplifies to $\mathbb{1}(y_{(k)i} = l)$. 

The gradients needed in Algorithm \ref{alg:1} can be expressed in terms of $\tilde{\mbP}$ and $\tilde{\mbC}.$ In particular, 
\begin{align*} \nabla_\beta \mathcal{F}_{0,0}(\mbalpha^t, \mbbeta, \mbgamma^t) &= \frac{1}{N} \sum_{k=1}^K \mbX_{(k)}^\top\left\{\tilde{\mbP}_{(k)}(\mbalpha^t, \mbbeta, \mbgamma^t_{(k)}) - \tilde{\mbC}_{(k)}(\mbalpha^t, \mbbeta, \mbgamma^t_{(k)})\right\},\\
\nabla_\alpha \mathcal{F}_{0,0}(\mbalpha, \mbbeta^{t+1}, \mbgamma^{t+1}) &= \frac{1}{N} \sum_{k=1}^K \left\{\tilde{\mbP}_{(k)}(\mbalpha, \mbbeta^{t+1}, \mbgamma_{(k)}^{t+1}) - \tilde{\mbC}_{(k)}(\mbalpha, \mbbeta^{t+1}, \mbgamma_{(k)}^{t+1})\right\}^\top \mathbf{ 1}_{n_k},\\
\nabla_{\gamma_{(k)}} \mathcal{F}_{0,0}(\mbalpha^t, \mbbeta^{t+1}, \mbgamma) &= \frac{1}{N} \mbZ_{(k)}^\top\left\{\tilde{\mbP}_{(k)}(\mbalpha^t, \mbbeta^{t+1}, \mbgamma_{(k)}) - \tilde{\mbC}_{(k)}(\mbalpha^t, \mbbeta^{t+1}, \mbgamma_{(k)})\right\},~~ k \in [K].
\end{align*}
Examining the form of these gradients, loosely speaking, we see our algorithm descends in the direction determined the correlation between the predictors and the difference between the unconditional and conditional estimated probabilities. The functions $\tilde{\mbP}$ and $\tilde{\mbC}$ are also used later when we apply our method to the motivating data analysis. 

In the Supplementary Material, we detail how we construct a set of candidate tuning parameters $(\lambda, \rho)$ yielding sparse fitted models. In brief, we use the KKT condition for \eqref{eq:estimator} to find a $\lambda$ yielding $\hat\mbbeta = \boldsymbol{0}$ and borrow an approach from \texttt{glmnet} for determining a reasonable set of values for $\rho$. 

\section{Simulation studies}\label{sec:simulation_studies}
We performed extensive numerical experiments to study how the sample size, number of predictors, similarity of categories, and the magnitude of batch effects affect the performance of various methods for estimating finest resolution cell type probabilities.

\subsection{Data generating models}
For each replication, we generated a total of $13$ datasets: six datasets with sample size $N / 6$ for fitting the model, six datasets with sample size $N / 6$ for validation, and one dataset with sample size $10^4$ for evaluating performance. We considered $N \in \left\{2400, 4800, 9600, 19200\right\}$ to reflect the large number of cells available in real datasets. We set the number of finest resolution categories to be fixed at $12$ ($\mathcal{C} = \{A_1, A_2, B_1, B_2, C_1, C_2, D_1, D_2, E_1, E_2, F_1, F_2\}$) and the binning functions fixed to have a structure inspired by the real data as shown by Figure 2. Specifically, in the real data, most cell types are observed at a coarse resolution in most datasets and at finest resolution in only a few datasets. Therefore, we chose to bin categories $A_1, A_2, B_1, B_2, C_1, C_2, D_1, D_2, E_1$, and $E_2$ into groups of two for Datasets 1--4. That is categories $A_1$ and $A_2$ are binned together, $B_1$ and $B_2$ are binned together, and so on. However, we set it so that these categories would be observed at the finest resolution in Datasets 5 and 6. Also, in the real data, some cell types are labeled at the finest resolution in all datasets (for example, CD14+ Monocytes and CD16+ Monocytes in Figure 2). Hence, we chose categories $F_1$ and $F_2$ to be observed at the finest resolution in all datasets. A graphical representation of these binning functions is shown in Figure B.1 of the Supplementary Material. The validation datasets, Datasets 7--12, are generated in the same way as Datasets 1--6. For the test dataset, all observations are observed at the finest resolution in order to fully evaluate parameter estimation.

In manual single-cell annotation, cell types are binned together due to their similar gene expression. We reflected this to varying extents in the structure of $\mbbeta^* \in \mathbb{R}^{p \times 12}$, where we consider $p \in \left\{250, 500, 1000, 2000\right\}$. We first randomly select $100$ of the $p$ rows to be nonzero in $\mbbeta^*$. Of these $100$ rows, we select $s$ many rows for which their coefficients are identical within the coarse groups described above, i.e. for these $s$ rows, the coefficients for category $A_1$ and $A_2$ are identical, coefficients for category $B_1$ and $B_2$ are identical, and so on. For the remaining $100 - s$ nonzero rows of $\mbbeta^*$, the coefficients for all categories are unrelated. We sample each of the nonzero distinct elements from a Normal$(0, 2)$ distribution. This structure to $\mbbeta^*$ controls the similarity of fine cell types within a coarse label. With $s = 0$, even though two categories may be binned together, they are unrelated and there is no true hierarchy of cell types. With larger $s$, fine categories within a coarse label are increasingly related, meaning there is true hierarchy to the cell type categories and cells are binned according to this hierarchy. We consider $s \in \left\{0, 20, 40, 60, 80\right\}$. 

Finally, to simulate the effect of batch effects in the predictors, we generated $\mbX_{(k)} = \tilde{\mbX}_{(k)} + \mbU_{(k)}$ where $\mbU_{(k)} = ( \mbu_{(k)1}, \dots, \mbu_{(k)n_k})^\top \in \mathbb{R}^{n_k \times p}$. Each row of $\tilde{\mbX}_{(k)}$ is independently simulated from a $p$-dimensional multivariate normal distribution with mean $0$ and AR(1) covariance matrix with lag $0.5$. We consider a simple model for the batch effect itself, in which the batch effect is identical for every observation within a batch. This may also be reasonable in the real data, as the presence of background contamination, also known as ambient RNA, is a common source of batch effects, and it may affect all cells within the experiment similarly (after normalization) \citep{young2020soupx}.  Therefore, we generate $\mbu_{(k)} \in \mathbb{R}^p$ as a realization from a $p$-dimensional mean zero multivariate normal distribution with covariance $\mbI_p$ and set $\mbU_{(k)i, :} = a \cdot \mbu_{(k)}$, where $a$ is a scalar chosen to control $b = \|\mbU_{(k)}\|_F / \|\tilde{\mbX}_{(k)}\|_F$. We consider $b \in \left\{0, 0.025, 0.05, 0.1, 0.2, 0.4\right\}$. The test dataset is observed with no batch effect, again in order to best evaluate parameter estimation.

\subsection{Competing methods}\label{subsec:comp_methods}
We first consider two variants of our method, \texttt{IBMR-int} and \texttt{IBMR-NG}. For \texttt{IBMR-int}, we set $\mbz_{(k)i} = 1$ for all  $i \in [n_k]$, $k \in [K]$, and fit the proposed model using \eqref{eq:estimator}. For \texttt{IBMR-NG}, we set $\mbgamma_{(k)} = 0$ for all $k \in [K]$, where ``NG'' stands for ``no Gamma'', and estimate only $\mbalpha^*$ and $\mbbeta^*$ using \eqref{eq:estimator}. This is a version of our method which ignores possible batches entirely.

We also consider two alternative methods, \texttt{subset} and \texttt{relabel}. For \texttt{subset}, we ``mix-and-match'' data from different datasets by subsetting each dataset to only the data that is annotated at the finest resolution and fit a model based on the stacked data. Specifically, define for $k \in [K]$, the set of indices in the $k$th dataset for which the outcome was observed at the finest resolution:
$
\mathcal{I}_k = \{ i : |g_k(y_{(k)i})| = 1 \}.
$
Then, we fit a group lasso-penalized multinomial logistic regression model using \eqref{eq:estimator}, but with $y_{(k)i}$ replaced with $g_k(y_{(k)i})$ for $k \in [K]$ and $i \in \mathcal{I}_k$, $\mathcal{C}_k$ replaced with $\mathcal{C}$ for $k \in [K]$, and $\mathcal{L}(\cdot, \cdot, \cdot)$ replaced with $ -(\sum_{k=1}^K |\mathcal{I}_k|)^{-1} \sum_{k=1}^k \sum_{i \in \mathcal{I}_k} l_{(k)i}(\cdot, \cdot, \cdot)$. However, because of potential confounding, we do not consider a batch effect (i.e., require $\mbgamma_{(k)} = \boldsymbol{0}$).
% Then, let $N = \sum_{k = 1}^{K}| \mathcal{I}_k|$ and $y \in \mathcal{C}^N$ be defined by 
% $
% y =(g_1(y_{(1)\mathcal{I}_1}), \dots, g_K (y_{(K)\mathcal{I}_K}) )^\top = (y_1, \dots, y_N)^\top
% $
% and $X \in \mathbb{R}^{N \times p}$ be defined by
% $
% X = \begin{pmatrix}
% X_{(1)\mathcal{I}_1, \cdot}' &
% \cdots &
% X_{(K)\mathcal{I}_K, \cdot}'
% \end{pmatrix}' = \left(x_1, \dots, x_N\right)'.
% $
% Then, we estimate
% $$
% \hat{\alpha}, \hat{\beta} = \argmin_{\alpha \in \mathbb{R}^{|\mathcal{C}|}, \beta \in \mathbb{R}^{p \times |\mathcal{C}|}} \left\{-\frac{1}{N} \sum_{i = 1}^{N} \sum_{l \in \mathcal{C}} \mathbb{1}(y_i = l) \log \left(\frac{{\rm exp}(\alpha_{l} + \beta_{l}'x_i)}{\sum_{v \in \mathcal{C}} {\rm exp}(\alpha_{v} + \beta_{v}'x_i)}\right) + \lambda \sum_{m = 1}^p \|\beta_{m, \cdot}\|_2  \right\}.
% $$ 
The model can thus be fit using existing software (e.g., \texttt{glmnet}), but since the objective function is identical to our method when using only subsetted data, we use our implementation for consistency in the algorithm and convergence criterion.

For the other method, \texttt{relabel}, we first obtain estimates of $(\mbalpha^*, \mbbeta^*)$ using \texttt{subset}, denoted $({\bar{\mbalpha}}^{\rm S}, {\bar{\mbbeta}}^{\rm S})$. Using these estimates, we can ``relabel'' our training data to have outcomes at the finest resolution by choosing the category with the highest conditional probability (as defined in \eqref{eq:Cmatrix})
$
\tilde{y}^{\rm S}_{(k)i} = \argmax_{l \in \mathcal{C}} [\tilde{\mbC}_{(k)}(\bar{\mbalpha}^{\rm S}, \bar{\mbbeta}^{\rm S}, \boldsymbol{0})]_{i, l}.
$
We then fit the multinomial logistic regression model to $\tilde{y}^{\rm S}$, treating these as the observed labels. To be clear, all the training responses $\tilde{y}^{\rm S}$ are (synthetically) at the finest resolution, so one fits \eqref{eq:estimator} but each $\mathcal{C}_k$ is replaced with $\mathcal{C}$.

Finally, we also consider oracle (\texttt{ORC}) versions of these methods, in which data at the finest resolution for all datasets is available. \texttt{IBMR-int-ORC} is the same as \texttt{IBMR-int}, with coarse resolution data replaced by the (otherwise unobserved) fine resolution data. By definition of \texttt{IBMR-NG}, \texttt{subset}, and \texttt{relabel}, when all the data is at the finest resolution, the estimators are equivalent to the standard group lasso penalized multinomial logistic regression model. Therefore, we name the oracle version of these estimators \texttt{GL-ORC}, where ``GL'' stands for ``group lasso.''

\subsection{Results}

\begin{figure}[!t]
\begin{center}
\includegraphics[width=\textwidth]{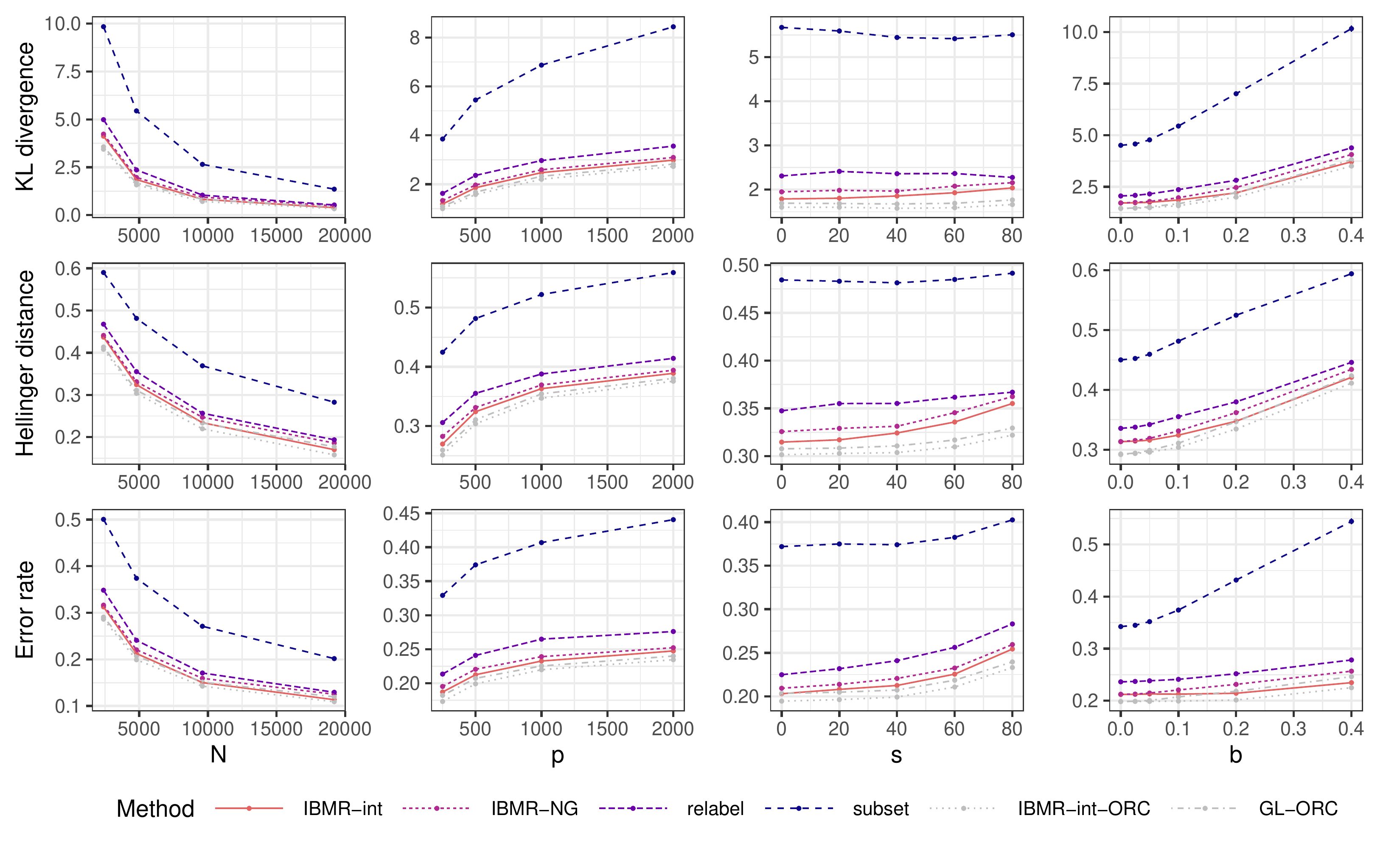}
\end{center}
\vspace{-20pt}
\caption{(top) Kullback-Leibler divergence, (middle) Hellinger distance, and (bottom) error rate for six competing methods with varying (left) $N$, the total sample size; (middle left) $p$, the total number of features; (middle right) $s$, the number of nonzero features which have shared coefficients for fine categories within a coarse label; and (right) $b$, the ratio of the norm of the batch effect and norm of the true predictors. Error bars denote the standard error for each method across 50 replicates. Throughout, the defaults are $N = 4800$, $p = 500$, $s = 40$, $b = 0.1$.} \label{fig:simulation}
\end{figure}

We present the complete simulation study results in Figure \ref{fig:simulation}. 
In the first column of Figure \ref{fig:simulation}, we present results with the total sample size $N \in \left\{2400, 4800, 9600, 19200\right\}$ varying, and $p = 500$, $s = 40$, $b = 0.1$ fixed. We see that with increasing sample size, the KL divergence, Hellinger distance, and error rates decrease for all methods, as expected. Of the non-oracle methods, for all sample sizes considered, \texttt{IBMR-int} and \texttt{IBMR-NG} perform the best and are much closer to the oracle methods in which all data is observed at the finest resolution, as compared to \texttt{relabel} and \texttt{subset}.  

In the second column of Figure \ref{fig:simulation}, we vary the total number of genes $p \in \left\{250, 500, 1000, 2000\right\}$ (all with $100$ nonzero rows of $\mbbeta^*$), with  $N = 4800$, $s = 40$, and $b = 0.1$ fixed. We see that with increasing number of genes, all performance metrics increase for all methods, as expected. Again, the \texttt{IBMR}-based methods are much closer to the oracle methods than \texttt{relabel} and especially \texttt{subset}. 

In the third column of Figure \ref{fig:simulation}, we vary the similarity of cell type categories within coarser groups by considering $s \in \left\{0, 20, 40, 60, 80\right\}$, the number of nonzero rows of $\mbbeta^*$ for which fine categories within a coarse label share coefficients. We fix $N = 4800$, $p = 500$, and $b = 0.1$. With $s$ increasing, fine categories within a coarse group become more similar, thus the Hellinger distance and error rates increase for all methods. This is because larger $s$ makes it more difficult to distinguish between the fine categories within a coarse group. KL divergence is relatively constant, but slightly increases for \texttt{IBMR}-based methods at $s$ increases. For all values of $s$, \texttt{IBMR}-based methods again perform more similar to the oracle methods than do \texttt{relabel} and \texttt{subset}.

For simulation results displayed in the last (rightmost) column of Figure \ref{fig:simulation}, we fixed $N = 4800$, $p = 500$, and $s = 40$ and varied the batch effect size by considering $b \in \{0, 0.025, 0.05, 0.1, 0.2, 0.4\}$. We see that with increasing batch effect, \texttt{IBMR-int} outperforms \texttt{IBMR-NG}, with the error rate of $\texttt{IBMR-int}$ staying relatively constant until $b = 0.2$. Of course $b = 0.2$ represents a quite large batch effect: in this situation the norm of the batch effect is, loosely speaking, 20 percent of the norm of the true gene expression. Again, \texttt{IBMR}-based methods are closest to oracle methods.

\section{Application to integrative cell type annotation}\label{sec:data_analysis}
%\subsection{Setup}
In this section, we apply our method to single-cell gene expression data from 10 publicly available peripheral blood mononuclear cells (PBMC) datasets with annotations at various resolutions and labels across datasets. These datasets can be downloaded in a standardized format as \texttt{Bioconductor SingleCellExperiment} objects from \url{https://github.com/keshav-motwani/AnnotatedPBMC}. Table \ref{table:datasets} lists the datasets used and the number of cell type labels per dataset. Table 3 gives the specific labels used in each dataset. The specifics of preprocessing of the data are described in Section A of the Supplementary Material.

\subsection{Comparison to \texttt{subset} and \texttt{relabel}}

In order to assess the performance of our method compared to competitors, we fit each method on eight datasets at a time, leaving out one validation dataset and one test dataset. In order to keep the binning functions the same across all train/validation/test splits, we kept the \texttt{hao_2020} dataset in the training set always because it had the finest resolution labels. We therefore defined the finest resolution categories across all datasets ($\mathcal{C}$) to be those used in the \texttt{hao_2020} dataset, and defined the binning functions ($f_k$) as graphically depicted in Figure \ref{fig:data_cell_types}. We evaluate performance over all 72 combinations of training/validation/test splits of eight training datasets (necessarily containing \texttt{hao_2020}), one validation dataset and one test dataset. We choose tuning parameters based on validation set negative log-likelihood, and measure performance using test set negative log-likelihood and error rate with the fitted parameters.

To reduce computational complexity, we perform screening on genes by ranking genes as described in Section A of the Supplementary Material, and select the first $p$ genes for each dataset. Also, for each training dataset, we sample $n_k$ cells using a weighted sampling procedure -- also described in Section A of the Supplementary Material -- in order to encourage oversampling extremely rare cell types and undersampling common cell types.

\begin{figure}[!t]
\begin{center}
\includegraphics[width=0.9\textwidth, page=2]{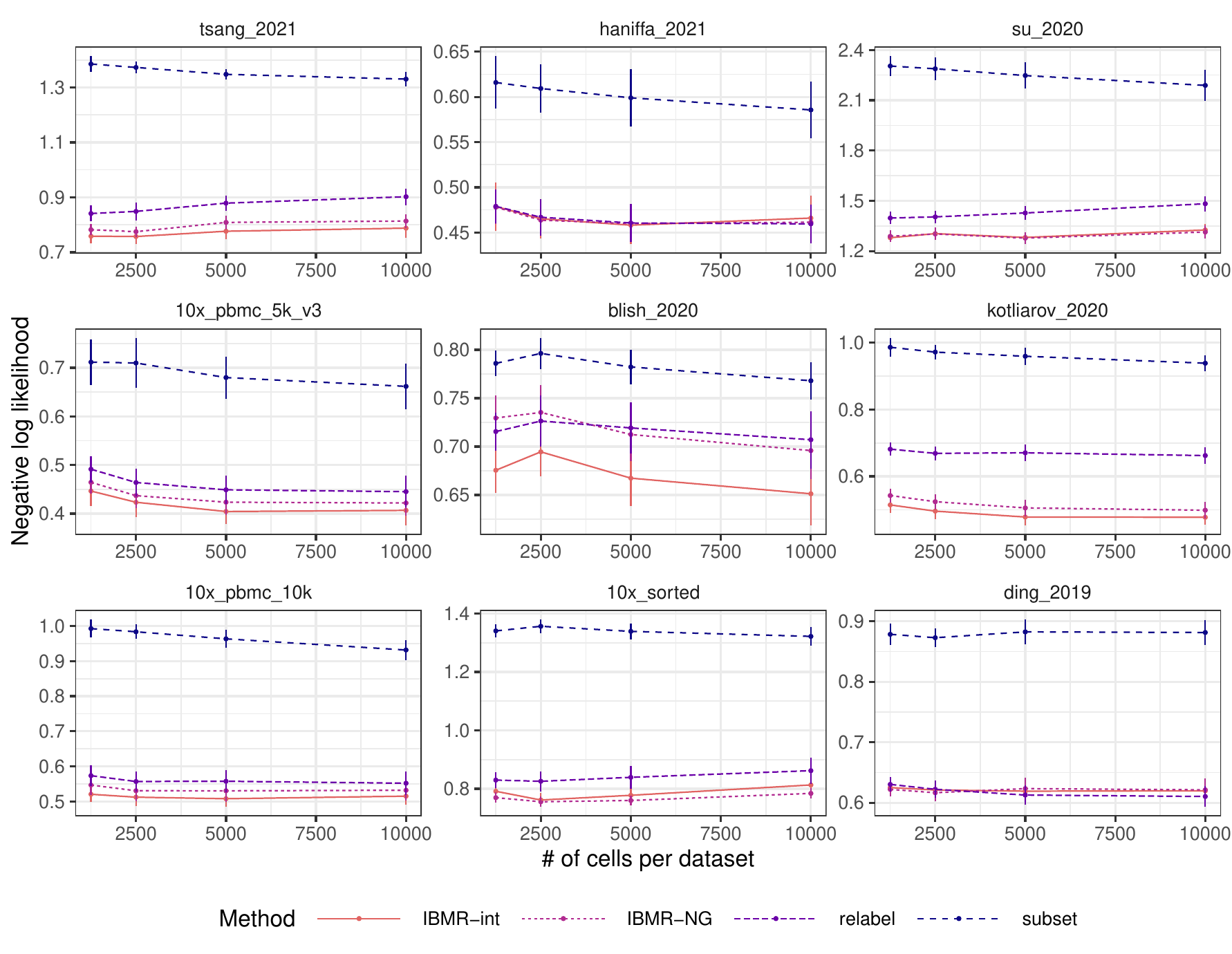}
\end{center}
\vspace{-25pt}
\caption{Negative log-likelihood for each method considered, for each test dataset (subplots), for varying numbers of cells per dataset used for fitting the model with the number of genes $p = 1000$ fixed. Points denote the average of the average negative log-likelihood across validation sets, for which each training/validation/test dataset combination had five replicates of different subsampled training datasets, and error bars show the standard error of averages across validation sets.} \label{fig:samplesizenll}
\end{figure}

We first assessed the test set negative log-likelihood of each of the non-oracle methods considered in the simulation study when varying the sample size per dataset $n_k \in \left\{1250, 2500, 5000, 10000 \right\}$ with the number of genes $p = 1000$ fixed. We repeat this five times, as the sampling of cells from each dataset is random. We then compute the negative log-likelihood for nine test datasets, each using one of the remaining eight datasets as a validation set, and the rest of the datasets as traning datasets, across five replicates. We first compute the average and standard error of the negative log likleihood across the five replicates for each train/validation/test dataset combination, and then summarize the results for each test dataset by taking the average and standard error of these averages across all of the train/validation dataset combinations considered. These summarized results per test dataset are shown in Figures \ref{fig:samplesizenll}, with the complete results for each validation and test dataset combination in Figure B.2 of the Supplementary Material. In general, the negative log-likelihood decreases or stays relatively constant with increasing sample size for all methods. \texttt{IBMR-int} tends to perform slightly better than \texttt{IBMR-NG} on some datasets, as fitting a batch-specific intercept term helps in these cases. In general, \texttt{IBMR}-based methods always do as well or better than \texttt{relabel}, and \texttt{subset} always performs the poorest. 

While the negative log-likelihood illustrates prediction performance in terms of estimated probabilities as a continuous value, it is more difficult to interpret than, say, classification error rate. For this reason, we also considered error rate, which is slightly more complicated to define in this setting. Specifically, in order to define an ``error,'' we must make predictions from the same set of labels used in the test dataset. We refer to these as ``coarse predictions'' and define them as follows.  Let $f_{\text{test}}: \mathcal{C} \to \mathcal{C}_{\text{test}}$ be the binning function for the test dataset labels, and $g_{\text{test}} = f_{\text{test}}^{-1}$ be the unbinning function as defined before. Because there may be labels in $\mathcal{C}_{\text{test}}$ which are bins of categories in $\mathcal{C}$ not observed in the test dataset in order to properly define the binning functions (named ``unobserved'' for example, as described earlier), we define $\ddot{\mathcal{C}}_{\text{test}}$ as follows:
$
\ddot{\mathcal{C}}_{\text{test}} = \{ j \in \mathcal{C}_{\text{test}}: \sum_{i = 1}^{n_{\text{test}}} \mathbb{1} \left( y_{(\text{test})i} = j \right) > 0 \}.
$
That is, it is a subset of $\mathcal{C}_{\text{test}}$ for which we actually observe cells annotated with that label.

\begin{figure}[!t]
\begin{center}
\includegraphics[width=0.9\textwidth, page=2]{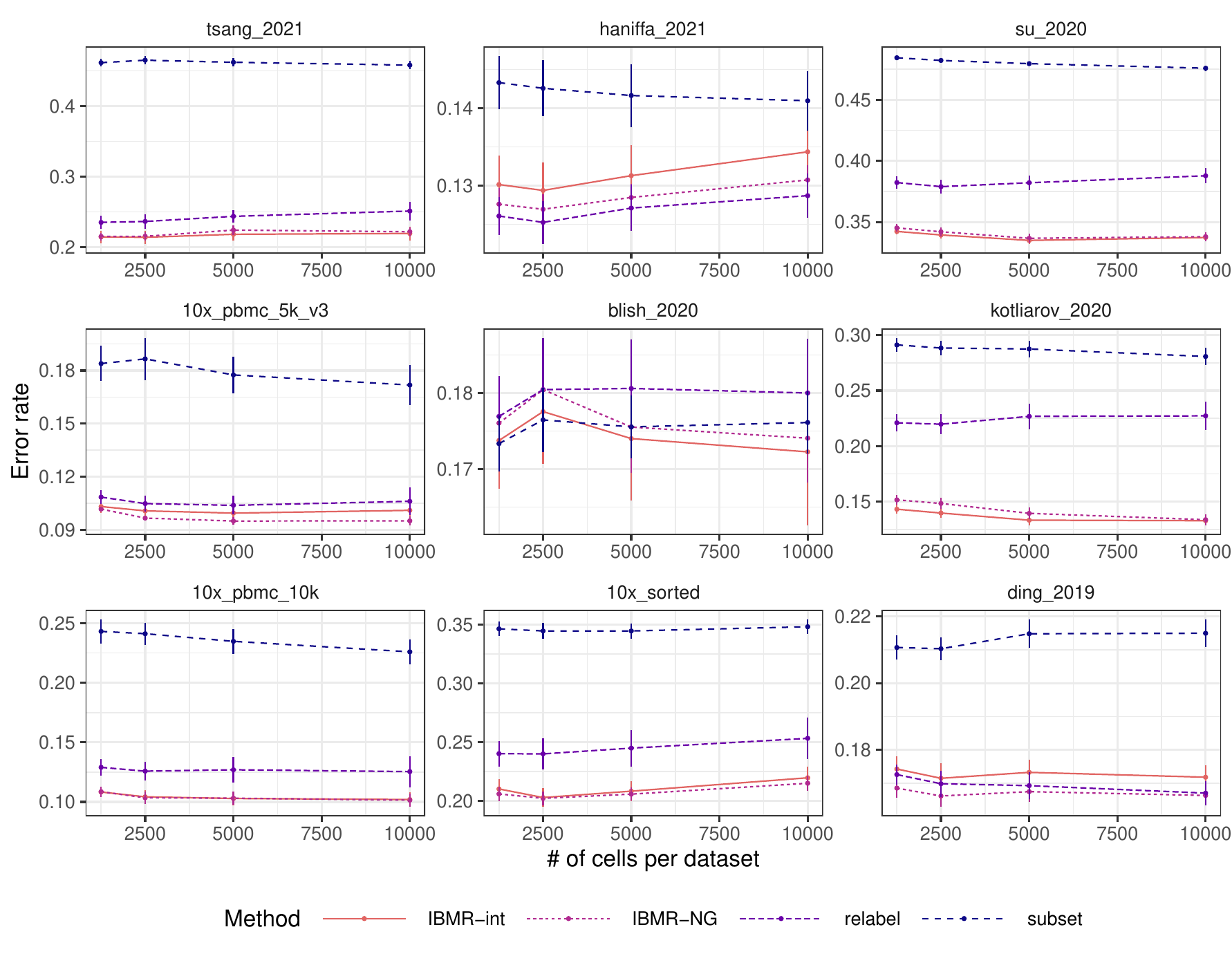}
\vspace{-25pt}
\end{center}
\caption{Error rate for each method considered, for each test dataset (subplots), for varying numbers of cells per dataset used for fitting the model with the number of genes $p = 1000$ fixed. Points denote the average of the average negative log-likelihood across validation sets, for which each training/validation/test dataset combination had five replicates of different subsampled training datasets, and error bars show the standard error of averages across validation sets.} \label{fig:samplesizeerror}
\end{figure}
We can then predict only within these labels to be consistent with the observed labels, which we call ``coarse predictions''. We have that the predicted probabilities at this coarse level are defined by 
$$
[\hat{\mbP}_{(\text{test})}]_{i, j} = \frac{\sum_{l \in g_{\text{test}}(j)} {\rm exp}(\hat{\mbalpha}_{l} + \mbx_{(\text{test})i}^\top \hat{\mbbeta}_{l} )}{\sum_{u \in \ddot{\mathcal{C}}_{\text{test}}} \sum_{v \in g_{\text{test}}(j)} {\rm exp}(\hat{\mbalpha}_{v} + \mbx_{(\text{test})i}^\top \hat{\mbbeta}_{v})}, ~~~ j \in \ddot{\mathcal{C}}_{\text{test}},
$$
and we then define the $i$th ``coarse prediction'' as 
$
\argmax_{j \in \ddot{\mathcal{C}}_{(\text{test})}} \{ [\hat{\mbP}_{(\text{test})}]_{i, j} \}.
$

The summarized error rate results per test dataset are shown in Figure \ref{fig:samplesizeerror} and complete results in Figure B.2 of the Supplementary Material. In general, there is increased variability in results across test datasets for error rate than for negative log-likelihood. Even so, in six out of nine test datasets considered, \texttt{IBMR-int} and \texttt{IBMR-NG} outperform \texttt{relabel} and \texttt{subset}, with \texttt{subset} resulting in error rates nearly double those of \texttt{IBMR-int}, \texttt{IBMR-NG}, and \texttt{relabel} in some cases.

% \begin{figure}[!ht]
% \centerfloat
% \begin{tabular}{C{.6\textwidth}C{.6\textwidth}}
% \subfigure [Negative log-likelihood] {
% \includegraphics[width=0.6\textwidth, page=2]{figures/application_figures_nll_1.pdf}
% } &
% \subfigure [Error rate] {
% \includegraphics[width=0.6\textwidth, page=2]{figures/application_figures_error_1.pdf}
% } \\
% \end{tabular}
% \caption{(a) Negative log-likelihood and (b) error rate for each method considered, for each test dataset (subplots), for varying numbers of cells per dataset used for fitting the model with the number of genes ($p$) fixed at $1000$. Points denote the average of the average negative log-likelihood/error rate across validation sets, for which each training/validation/test dataset combination had five replicates of different subsampled training datasets, and error bars show the standard error of averages across validation sets.} \label{fig:samplesize}
% \end{figure}

We next performed a similar experiment: with the sample size per dataset $n_k = 5000$ fixed, we  varied the number of predictors $p \in \left\{250, 500, 1000, 2000\right\}$. Once again, we adopted the same setup for training/validation/test splits, and 5 replicates per split to account for subsampling variability. The summarized results per test dataset are shown in the Supplementary Material, in Figures B.4 (negative log-likelihood) and B.5 (error rate), with the complete results for each validation and test dataset combination in Figures B.6 (negative log-likelihood) and B.7 (error rate). Overall, we see that accounting for the batch effect with \texttt{IBMR-int} usually improves upon \texttt{IBMR-NG}, with \texttt{relabel} generally falling behind \texttt{IBMR}-based methods. The method \texttt{subset} consistently performs poorly compared to the other methods for all datasets. 

% \begin{figure}[!ht]{}
% \begin{center}
% \includegraphics[width=\textwidth, page=1]{figures/application_figures_nll_1.pdf}
% \end{center}
% \caption{Same as the previous figure, but now varying numbers of genes used for fitting the model with the number of cells per dataset $n_k = 5000$ fixed.} \label{fig:genes}
% \end{figure}
\begin{figure}[t]
\begin{center}
\includegraphics[width=\textwidth, page=1]{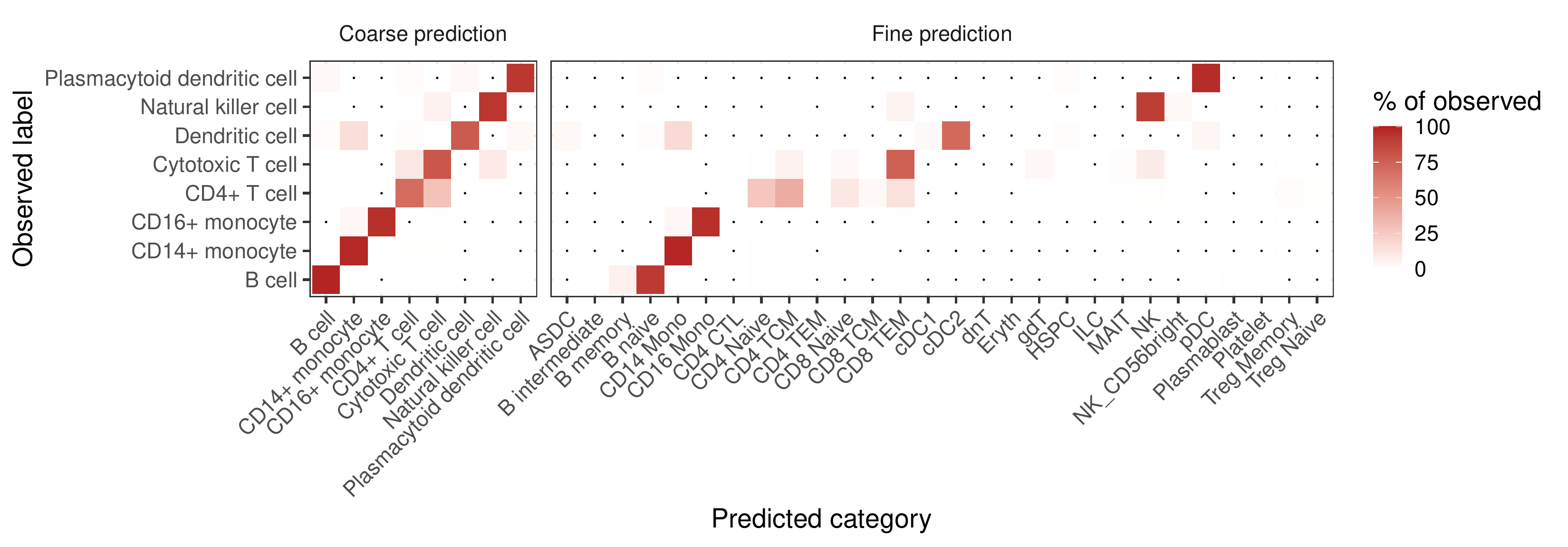}
\end{center}
\vspace{-25pt}
\caption{Heatmap showing the percentage of cells in (left) coarse and (right) fine predicted categories for each observed label in the \texttt{ding_2019} dataset. Dots indicate that exactly 0 cells are in that combination of observed label and prediction.} \label{fig:heatmapfine}
\end{figure}

\subsection{Annotating or refining cell type labels on new datasets}

In this section, we use our fitted model to annotate and refine cell type labels on a new dataset. For this, we turn our attention to the \texttt{IBMR-int} model fit in the last section with \texttt{tsang_2021} as the validation set and \texttt{ding_2019} as the test dataset, for the first replicate of the experiment with $n_k = 10000$ and $p = 1000$. We choose \texttt{tsang_2021} to be the validation set because it has the finest annotations over all validation sets considered and we chose to predict on \texttt{ding_2019} because it has the most coarse annotations.

\begin{figure}[t]
\begin{center}
\includegraphics[width=\textwidth, page=1]{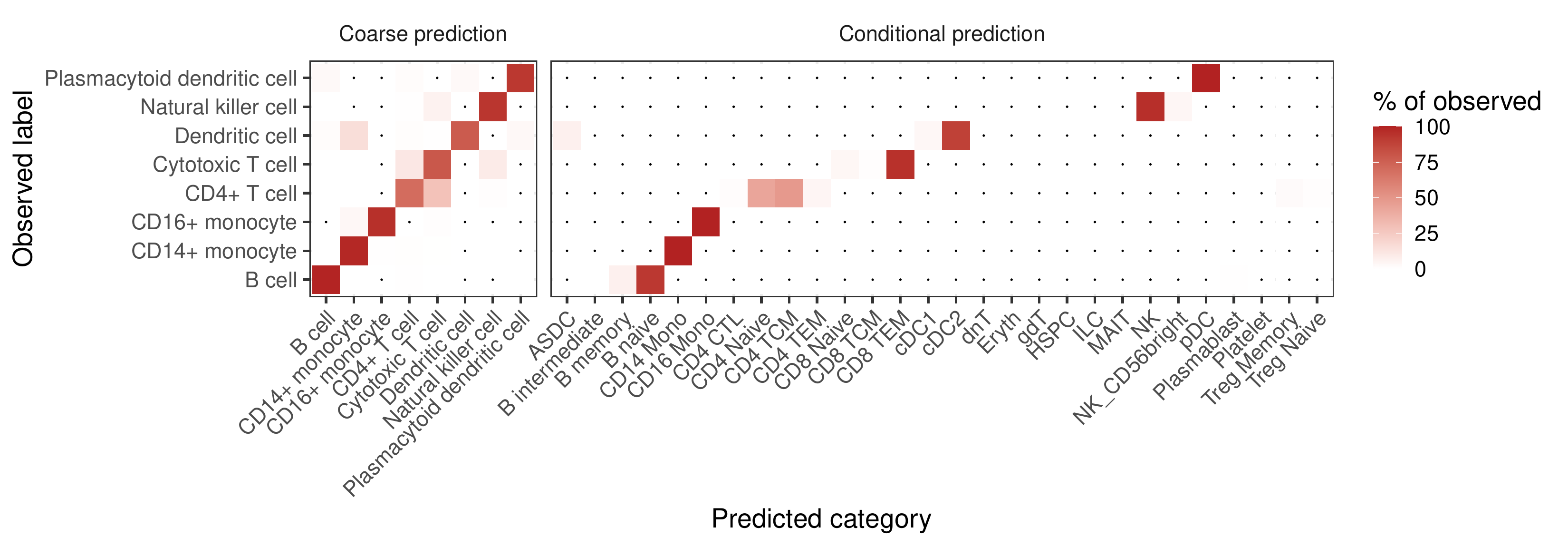}
\end{center}
\vspace{-25pt}
\caption{Heatmap showing the percentage of cells in (left) coarse and (right) conditional predicted categories for each observed label in the \texttt{ding_2019} dataset. Dots indicate that exactly 0 cells are in that combination of observed label and prediction. Note that the difference between Figure \ref{fig:heatmapfine} and this figure is that the (right) panel is showing conditional predictions rather than fine predictions.} \label{fig:heatmapcond}
\end{figure}

There are three types of predictions we may consider: (i) a prediction of the finest resolution categories based on our model, which was the primary motivation; (ii) if we have already observed coarse annotations on a dataset, we can make predictions of the finest resolution categories, conditional on already observed coarse label; or (iii) coarse predictions as described in the previous section for performance evaluation. Note that (ii) is especially useful when only coarse labels are used in annotating a dataset initially, but more refined annotations are desired for downstream analyses. 

In the case where we are simply interested in predicting fine resolution categories on a dataset with only gene expression observed, (i), we define the $i$th ``fine prediction'' as 
$
\argmax_{l \in \mathcal{C}} \{ [\tilde{\mbP}_{(\text{test})}(\hat{\mbalpha}, \hat{\mbbeta}, \boldsymbol{0})]_{i, l} \}
$
where $\tilde{\mbP}_{(\text{test})}$ is defined as in \eqref{eq:Pmatrix}.
Alternatively, if we want to predict fine resolution categories, but have observed both gene expression and coarse resolution annotations, we can condition on the coarse label and obtain conditional predictions, i.e., prediction of type (ii). In effect, this refines the existing annotations based on the fitted model and provides more detailed annotations. In this case, we define the ``conditional predictions'' as 
$
\argmax_{l \in \mathcal{C}} \{ [\tilde{\mbC}_{(\text{test})}(\hat{\mbalpha}, \hat{\mbbeta}, \boldsymbol{0})]_{i,l} \}
$
where $\tilde{\mbC}_{(\text{test})}$ is defined as in \eqref{eq:Cmatrix}. Note that if an observation already has a fine label, then the label will not change by the definition of the conditional probabilities, so there will not be contradictory results.

In Figure \ref{fig:heatmapfine}, we show the coarse predictions and fine predictions, and the percentage of each of the observed labels which are predicted as a given category for these two types of predictions. The model does very well at predicting at the coarse level, with few predictions in the off-diagonal elements of the heatmap. The fine predictions generally agree very well with the coarse observed annotations, while giving additional information. In Figure \ref{fig:heatmapcond}, we once again show the same coarse predictions as a reference, and also take advantage of the already coarsely labeled data to provide predictions conditional on the observed coarse annotations. These conditional predictions only split up an observed label into finer categories by definition, so they provide additional detail and do not ever contradict the initial coarse annotations.

In order to showcase interpretability of the model coefficients for this same fitted model considered above, we display the genes corresponding to the top 10 standardized coefficients per finest resolution category in Table \ref{table:genes}. Many of these genes overlap with commonly used marker genes for these cell types, as shown in bold in \ref{table:genes}, based on marker genes by \citet{hao2020integrated} for these categories. Note that these marker genes were defined by \citet{hao2020integrated} only on the \texttt{hao_2020} dataset, and were the result of performing hypothesis testing on the gene expression of a particular gene within cells of a category compared to all other cells, so these same genes may not be optimal for classification purposes.

% latex table generated in R 4.0.5 by xtable 1.8-4 package
% Mon Nov  1 18:57:34 2021

\section{Discussion}
In this article, we proposed a new method for integrating multiple datasets where observation labels are available at different resolutions. Overall, \texttt{IBMR}-based methods outperformed other competitors under all simulation settings, and generally performed better in the application to single-cell genomics data, with \texttt{relabel} only having close performance in a small number of cases. However, there are additional aspects of the methods to be considered in terms of performance and practical usage. Specifically, while \texttt{relabel} could be considered a two-step ``approximation'' to \texttt{IBMR-NG}, it involves two tuning parameters, and for each tuning parameter combination, two optimization problems must be solved. Each of these problems are similar in complexity to \texttt{IBMR-NG}. Therefore, \texttt{IBMR-NG}, which only involves one tuning parameter and one optimization problem, is arguably preferable to \texttt{relabel} as it is faster and, in general, more accurately estimates test set probabilities. Additionally, in the real data and even under large batch effects in simulations, \texttt{IBMR-NG} performs similarly to \texttt{IBMR-int}, which also involves two tuning parameters. Therefore, it may be reasonable to use \texttt{IBMR-NG} as an approximation to \texttt{IBMR-int} to further reduce computing times. These latter findings cohere with those in \citet{ma2021evaluation}, who found adjusting for batch effects did not have a substantial impact on cell type prediction.

\begin{table}[t]
\centering
\resizebox{\columnwidth}{!}{%
\begin{tabular}{r| llllllllll}
\toprule
Cell type & \multicolumn{10}{c}{Genes}\\
\midrule
ASDC & TCF4 & SOX4 & \textbf{PPP1R14A} & HLA-DRA & GPR183 & S100A4 & IRF8 & CD74 & ITM2C & S100A10 \\ 
  B intermediate & \textbf{MS4A1} & \textbf{CD79A} & \textbf{BANK1} & \textbf{RALGPS2} & GPR183 & HLA-DRA & TCF4 & MALAT1 & \textbf{TNFRSF13B} & TNFRSF13C \\ 
  B memory & \textbf{MS4A1} & HLA-DRA & \textbf{BANK1} & LTB & \textbf{CD79A} & ITGB1 & TNFRSF13B & CD74 & MALAT1 & \textbf{TNFRSF13C} \\ 
  B naive & \textbf{MS4A1} & \textbf{CD79A} & CD74 & HLA-DRA & \textbf{TCL1A} & LTB & FCGR3A & BANK1 & LINC00926 & \textbf{YBX3} \\ 
  CD14 Mono & \textbf{LYZ} & \textbf{S100A8} & TYROBP & HLA-DRA & \textbf{FCN1} & \textbf{CD14} & FTL & PSAP & AIF1 & HLA-DRB1 \\ 
  CD16 Mono & \textbf{FCGR3A} & \textbf{AIF1} & \textbf{MS4A7} & \textbf{LST1} & \textbf{CDKN1C} & TYROBP & PSAP & NAP1L1 & S100A4 & \textbf{IFITM3} \\ 
  CD4 CTL & CCL5 & \textbf{GNLY} & IL7R & \textbf{NKG7} & S100A4 & \textbf{ITGB1} & CD3G & \textbf{IL32} & MALAT1 & CD3D \\ 
  CD4 Naive & LTB & \textbf{CCR7} & CD3D & MALAT1 & CD3E & \textbf{NOSIP} & CD7 & NKG7 & \textbf{FHIT} & \textbf{LDHB} \\ 
  CD4 TCM & S100A4 & CD3D & \textbf{LTB} & \textbf{IL7R} & CD52 & ANXA1 & \textbf{ITGB1} & \textbf{IL32} & S100A11 & CD40LG \\ 
  CD4 TEM & \textbf{IL7R} & \textbf{GZMK} & \textbf{CCL5} & \textbf{KLRB1} & \textbf{IL32} & \textbf{LTB} & GPR183 & CD3G & S100A4 & MALAT1 \\ 
  CD8 Naive & \textbf{CD8B} & \textbf{CD8A} & CTSW & CD3D & \textbf{S100B} & MALAT1 & FCGR3A & AIF1 & CD7 & HCST \\ 
  CD8 TCM & \textbf{CD8B} & \textbf{CD8A} & \textbf{IL7R} & CCL5 & IL32 & LTB & S100A4 & ITGB1 & \textbf{ANXA1} & CTSW \\ 
  CD8 TEM & \textbf{CCL5} & \textbf{CD8A} & \textbf{CD8B} & \textbf{GZMK} & \textbf{NKG7} & CD3D & CTSW & IL32 & MALAT1 & \textbf{GZMH} \\ 
  cDC1 & HLA-DRA & \textbf{CADM1} & CD74 & IRF8 & HLA-DPB1 & HLA-DPA1 & LYZ & ID2 & S100A10 & HLA-DRB1 \\ 
  cDC2 & CD74 & \textbf{FCER1A} & HLA-DRA & \textbf{CD1C} & HLA-DPA1 & TYROBP & HLA-DPB1 & VIM & S100A10 & CST3 \\ 
  dnT & \textbf{GZMK} & \textbf{NUCB2} & GPR183 & CD8B & MALAT1 & CD3D & CD3G & HBB & \textbf{FXYD2} & CCR7 \\ 
  Eryth & HBB & CD8B & CD8A & FCGR3A & IL7R & MS4A1 & AHNAK & PSAP & DUSP1 & TNFAIP3 \\ 
  gdT & CCL5 & IL7R & \textbf{KLRD1} & CD3D & \textbf{KLRC1} & IL32 & CD3G & KLRB1 & \textbf{NKG7} & RTKN2 \\ 
  HSPC & \textbf{SPINK2} & \textbf{PRSS57} & SOX4 & AIF1 & RPS20 & HLA-DRA & \textbf{CYTL1} & PPBP & CD79A & LST1 \\ 
  ILC & \textbf{KLRB1} & IL7R & ITGB1 & \textbf{TNFRSF18} & GPR183 & \textbf{TNFRSF4} & LTB & IL2RA & MALAT1 & SPINK2 \\ 
  MAIT & \textbf{KLRB1} & \textbf{GZMK} & \textbf{IL7R} & CD8A & CD8B & CCL5 & S100A4 & LTB & \textbf{NKG7} & \textbf{NCR3} \\ 
  NK & \textbf{GNLY} & FCGR3A & \textbf{TYROBP} & \textbf{PRF1} & CTSW & \textbf{NKG7} & KLRB1 & KLRD1 & \textbf{KLRF1} & CD247 \\ 
  NK_CD56bright & GNLY & GZMK & \textbf{XCL1} & CTSW & \textbf{KLRC1} & KLRD1 & TYROBP & KLRB1 & \textbf{XCL2} & NKG7 \\ 
  pDC & TCF4 & \textbf{ITM2C} & \textbf{MZB1} & \textbf{SERPINF1} & CD74 & IRF8 & \textbf{PLD4} & HLA-DRA & TCL1A & GPR183 \\ 
  Plasmablast & \textbf{MZB1} & CD79A & ITM2C & ITGB1 & \textbf{TNFRSF13B} & PRDM1 & \textbf{CPNE5} & AQP3 & \textbf{POU2AF1} & TCF4 \\ 
  Platelet & \textbf{PPBP} & \textbf{TUBB1} & CD8B & SPARC & \textbf{NRGN} & HBB & CCL5 & IL7R & CD8A & ITGB1 \\ 
  Treg Memory & \textbf{RTKN2} & IL32 & \textbf{TIGIT} & \textbf{CTLA4} & \textbf{FOXP3} & \textbf{S100A4} & \textbf{IKZF2} & IL2RA & ITGB1 & LTB \\ 
  Treg Naive & IL32 & \textbf{RTKN2} & CD3D & CD3E & \textbf{IL2RA} & LTB & \textbf{FOXP3} & DUSP1 & CTLA4 & IKZF2 \\ 
   \bottomrule
\end{tabular}
}
\caption{Top 10 genes with largest standardized coefficients for each of the finest resolution categories (rows). These genes align with commonly used marker genes (bolded) for manually annotating cell types based on \citet{hao2020integrated}.} \label{table:genes}
\end{table}

There are multiple directions for future research. First, we have assumed a multinomial logistic regression model. Instead, it may be preferable to use a semiparametric or nonparameteric approach for modeling the probabilities \eqref{eq:probabilities}. For example, the application of random forests to this context may perform well. Second, our method did not exploit the similarity of cell types within a coarse category in any way. For example, in Section \ref{sec:simulation_studies} we generated data such that coefficient vectors for two cell types belonging to a coarse category were more similar compared to cell types which did not belong to a shared coarse category. We are currently developing an extension of our method which can exploit this feature. 

\subsection*{Acknowledgements}
Keshav Motwani's research was supported by the Goldwater Foundation as well as the University Scholars Program at the University of Florida. Aaron J. Molstad's research was supported by National Science Foundation grant DMS-2113589. 

\bibliography{rna2prot}

\begin{thebibliography}{}

\bibitem[10x Genomics, 2018]{10x_pbmc_10k}
10x Genomics (2018).
\newblock 10k {PBMC}s from a healthy donor - gene expression and cell surface
  protein.
\newblock
  \url{https://support.10xgenomics.com/single-cell-gene-expression/datasets/3.0.0/pbmc_10k_protein_v3}.

\bibitem[10x Genomics, 2019]{10x_pbmc_5k_v3}
10x Genomics (2019).
\newblock 5k {P}eripheral blood mononuclear cells ({PBMC}s) from a healthy
  donor with cell surface proteins (v3 chemistry).
\newblock
  \url{https://support.10xgenomics.com/single-cell-gene-expression/datasets/3.0.2/5k_pbmc_protein_v3}.

\bibitem[Abdelaal et~al., 2019]{abdelaal2019comparison}
Abdelaal, T., Michielsen, L., Cats, D., Hoogduin, D., Mei, H., Reinders, M.~J.,
  and Mahfouz, A. (2019).
\newblock A comparison of automatic cell identification methods for single-cell
  {RNA} sequencing data.
\newblock {\em Genome Biology}, 20(1):194.

\bibitem[Amezquita et~al., 2020]{amezquita2020orchestrating}
Amezquita, R.~A., Lun, A.~T., Becht, E., Carey, V.~J., Carpp, L.~N.,
  Geistlinger, L., Marini, F., Rue-Albrecht, K., Risso, D., and Soneson, C.
  (2020).
\newblock Orchestrating single-cell analysis with bioconductor.
\newblock {\em Nature Methods}, 17(2):137--145.

\bibitem[Conde et~al., 2021]{conde2021cross}
Conde, C.~D., Gomes, T., Jarvis, L.~B., Xu, C., Howlett, S., Rainbow, D.,
  Suchanek, O., King, H., Mamanova, L., and Polanski, K. (2021).
\newblock Cross-tissue immune cell analysis reveals tissue-specific adaptations
  and clonal architecture across the human body.
\newblock {\em bioRxiv}.

\bibitem[Ding et~al., 2019]{ding2019systematic}
Ding, J., Adiconis, X., Simmons, S.~K., Kowalczyk, M.~S., Hession, C.~C.,
  Marjanovic, N.~D., Hughes, T.~K., Wadsworth, M.~H., Burks, T., and Nguyen,
  L.~T. (2019).
\newblock Systematic comparative analysis of single cell rna-sequencing
  methods.
\newblock {\em BioRxiv}, page 632216.

\bibitem[Haghverdi et~al., 2018]{haghverdi2018batch}
Haghverdi, L., Lun, A.~T., Morgan, M.~D., and Marioni, J.~C. (2018).
\newblock Batch effects in single-cell {RNA}-sequencing data are corrected by
  matching mutual nearest neighbors.
\newblock {\em Nature Biotechnology}, 36(5):421--427.

\bibitem[Hao et~al., 2020]{hao2020integrated}
Hao, Y., Hao, S., Andersen-Nissen, E., Mauck, W.~M., Zheng, S., Butler, A.,
  Lee, M.~J., Wilk, A.~J., Darby, C., and Zagar, M. (2020).
\newblock Integrated analysis of multimodal single-cell data.
\newblock {\em bioRxiv}.

\bibitem[Hie et~al., 2019]{hie2019efficient}
Hie, B., Bryson, B., and Berger, B. (2019).
\newblock Efficient integration of heterogeneous single-cell transcriptomes
  using {S}canorama.
\newblock {\em Nature Biotechnology}, 37(6):685--691.

\bibitem[Huang et~al., 2017]{huang2017promoting}
Huang, Y., Zhang, Q., Zhang, S., Huang, J., and Ma, S. (2017).
\newblock Promoting similarity of sparsity structures in integrative analysis
  with penalization.
\newblock {\em Journal of the American Statistical Association},
  112(517):342--350.

\bibitem[Korsunsky et~al., 2019]{korsunsky2019fast}
Korsunsky, I., Millard, N., Fan, J., Slowikowski, K., Zhang, F., Wei, K.,
  Baglaenko, Y., Brenner, M., Loh, P.-r., and Raychaudhuri, S. (2019).
\newblock Fast, sensitive and accurate integration of single-cell data with
  {H}armony.
\newblock {\em Nature Methods}, 16(12):1289--1296.

\bibitem[Kotliarov et~al., 2020]{kotliarov2020broad}
Kotliarov, Y., Sparks, R., Martins, A.~J., Mul{\`e}, M.~P., Lu, Y., Goswami,
  M., Kardava, L., Banchereau, R., Pascual, V., and Biancotto, A. (2020).
\newblock Broad immune activation underlies shared set point signatures for
  vaccine responsiveness in healthy individuals and disease activity in
  patients with lupus.
\newblock {\em Nature Medicine}, 26(4):618--629.

\bibitem[Lange, 2016]{lange2016mm}
Lange, K. (2016).
\newblock {\em {MM} optimization algorithms}.
\newblock SIAM.

\bibitem[Liu et~al., 2021]{tsangliu2021time}
Liu, C., Martins, A.~J., Lau, W.~W., Rachmaninoff, N., Chen, J., Imberti, L.,
  Mostaghimi, D., Fink, D.~L., Burbelo, P.~D., and Dobbs, K. (2021).
\newblock Time-resolved systems immunology reveals a late juncture linked to
  fatal {COVID}-19.
\newblock {\em Cell}, 184(7):1836--1857.

\bibitem[Ma et~al., 2021]{ma2021evaluation}
Ma, W., Su, K., and Wu, H. (2021).
\newblock Evaluation of some aspects in supervised cell type identification for
  single-cell {RNA}-seq: classifier, feature selection, and reference
  construction.
\newblock {\em Genome Biology}, 22(1):1--23.

\bibitem[Molstad and Patra, 2021]{molstad2021dimension}
Molstad, A.~J. and Patra, R.~K. (2021).
\newblock Dimension reduction for integrative survival analysis.
\newblock {\em arXiv preprint arXiv:2108.02143}.

\bibitem[Molstad and Rothman, 2021]{molstad2021likelihood}
Molstad, A.~J. and Rothman, A.~J. (2021).
\newblock A likelihood-based approach for multivariate categorical response
  regression in high dimensions.
\newblock {\em Journal of the American Statistical Association}.

\bibitem[Obozinski et~al., 2011]{obozinski2011support}
Obozinski, G., Wainwright, M.~J., and Jordan, M.~I. (2011).
\newblock Support union recovery in high-dimensional multivariate regression.
\newblock {\em The Annals of Statistics}, 39(1):1--47.

\bibitem[Parikh and Boyd, 2014]{parikh2014proximal}
Parikh, N. and Boyd, S. (2014).
\newblock Proximal algorithms.
\newblock {\em Foundations and Trends in Optimization}, 1(3):127--239.

\bibitem[Pasquini et~al., 2021]{PASQUINI2021961}
Pasquini, G., {Rojo Arias}, J.~E., Schäfer, P., and Busskamp, V. (2021).
\newblock Automated methods for cell type annotation on sc{RNA}-seq data.
\newblock {\em Computational and Structural Biotechnology Journal},
  19:961--969.

\bibitem[Polson et~al., 2015]{polson2015proximal}
Polson, N.~G., Scott, J.~G., and Willard, B.~T. (2015).
\newblock Proximal algorithms in statistics and machine learning.
\newblock {\em Statistical Science}, 30(4):559--581.

\bibitem[Schaum et~al., 2018]{tabula2018single}
Schaum, N., Karkanias, J., Neff, N.~F., May, A.~P., Quake, S.~R., Wyss-Coray,
  T., Darmanis, S., Batson, J., Botvinnik, O., and Chen, M.~B. (2018).
\newblock Single-cell transcriptomics of 20 mouse organs creates a tabula
  muris: The tabula muris consortium.
\newblock {\em Nature}, 562(7727):367.

\bibitem[Shasha et~al., 2021]{shasha2021superscan}
Shasha, C., Tian, Y., Mair, F., Miller, H.~E., and Gottardo, R. (2021).
\newblock Superscan: Supervised single-cell annotation.
\newblock {\em bioRxiv}.

\bibitem[Simon et~al., 2013]{simon2013sparse}
Simon, N., Friedman, J., Hastie, T., and Tibshirani, R. (2013).
\newblock A sparse-group lasso.
\newblock {\em Journal of Computational and Graphical Statistics},
  22(2):231--245.

\bibitem[Stephenson et~al., 2021]{haniffastephenson2021single}
Stephenson, E., Reynolds, G., Botting, R.~A., Calero-Nieto, F.~J., Morgan,
  M.~D., Tuong, Z.~K., Bach, K., Sungnak, W., Worlock, K.~B., and Yoshida, M.
  (2021).
\newblock Single-cell multi-omics analysis of the immune response in
  {COVID}-19.
\newblock {\em Nature Medicine}, 27(5):904--916.

\bibitem[Su et~al., 2020]{su2020multi}
Su, Y., Chen, D., Yuan, D., Lausted, C., Choi, J., Dai, C.~L., Voillet, V.,
  Duvvuri, V.~R., Scherler, K., and Troisch, P. (2020).
\newblock Multi-omics resolves a sharp disease-state shift between mild and
  moderate {COVID}-19.
\newblock {\em Cell}, 183(6):1479--1495.

\bibitem[Ventz et~al., 2021]{ventz2021integration}
Ventz, S., Mazumder, R., and Trippa, L. (2021).
\newblock Integration of survival data from multiple studies.
\newblock {\em Biometrics}.

\bibitem[Wilk et~al., 2020]{blishwilk2020single}
Wilk, A.~J., Rustagi, A., Zhao, N.~Q., Roque, J., Mart{\'\i}nez-Col{\'o}n,
  G.~J., McKechnie, J.~L., Ivison, G.~T., Ranganath, T., Vergara, R., and
  Hollis, T. (2020).
\newblock A single-cell atlas of the peripheral immune response in patients
  with severe {COVID}-19.
\newblock {\em Nature Medicine}, 26(7):1070--1076.

\bibitem[Wolf et~al., 2018]{wolf2018scanpy}
Wolf, F.~A., Angerer, P., and Theis, F.~J. (2018).
\newblock Scanpy: large-scale single-cell gene expression data analysis.
\newblock {\em Genome Biology}, 19(1):1--5.

\bibitem[Xu and Yin, 2017]{xu2017globally}
Xu, Y. and Yin, W. (2017).
\newblock A globally convergent algorithm for nonconvex optimization based on
  block coordinate update.
\newblock {\em Journal of Scientific Computing}, 72(2):700--734.

\bibitem[Young and Behjati, 2020]{young2020soupx}
Young, M.~D. and Behjati, S. (2020).
\newblock {Soup{X} removes ambient {RNA} contamination from droplet-based
  single-cell {RNA} sequencing data}.
\newblock {\em GigaScience}, 9(12).
\newblock giaa151.

\bibitem[Yuan and Lin, 2006]{yuan2006model}
Yuan, M. and Lin, Y. (2006).
\newblock Model selection and estimation in regression with grouped variables.
\newblock {\em Journal of the Royal Statistical Society: Series B (Statistical
  Methodology)}, 68(1):49--67.

\bibitem[Zhao et~al., 2015]{zhao2015integrative}
Zhao, Q., Shi, X., Huang, J., Liu, J., Li, Y., and Ma, S. (2015).
\newblock Integrative analysis of ‘-omics’ data using penalty functions.
\newblock {\em Wiley Interdisciplinary Reviews: Computational Statistics},
  7(1):99--108.

\bibitem[Zheng et~al., 2017]{zheng2017massively}
Zheng, G.~X., Terry, J.~M., Belgrader, P., Ryvkin, P., Bent, Z.~W., Wilson, R.,
  Ziraldo, S.~B., Wheeler, T.~D., McDermott, G.~P., and Zhu, J. (2017).
\newblock Massively parallel digital transcriptional profiling of single cells.
\newblock {\em Nature Communications}, 8(1):1--12.

\end{thebibliography}

\clearpage

\end{document}